\documentstyle[epsfig]{mn}

\newif\ifAMStwofonts

\ifoldfss
  \ifCUPmtlplainloaded \else
    \NewTextAlphabet{textbfit} {cmbxti10} {}
    \NewTextAlphabet{textbfss} {cmssbx10} {}
    \NewMathAlphabet{mathbfit} {cmbxti10} {} 
    \NewMathAlphabet{mathbfss} {cmssbx10} {} 
  \fi
  \ifAMStwofonts
    \ifCUPmtlplainloaded \else
      \NewSymbolFont{upmath} {eurm10}
      \NewSymbolFont{AMSa} {msam10}
      \NewMathSymbol{\upi}     {0}{upmath}{19}
      \NewMathSymbol{\umu}     {0}{upmath}{16}
      \NewMathSymbol{\upartial}{0}{upmath}{40}
      \NewMathSymbol{\leqslant}{3}{AMSa}{36}
      \NewMathSymbol{\geqslant}{3}{AMSa}{3E}

      \let\leq=\leqslant 
      \let\geq=\geqslant \let\ge=\geqslant
    \fi
  \fi
\fi 

\ifnfssone
  \newmathalphabet{\mathit}
  \addtoversion{normal}{\mathit}{cmr}{m}{it}
  \addtoversion{bold}{\mathit}{cmr}{bx}{it}
  \newmathalphabet{\mathbfit} 
  \addtoversion{normal}{\mathbfit}{cmr}{bx}{it}
  \addtoversion{bold}{\mathbfit}{cmr}{bx}{it}
  \newmathalphabet{\mathbfss} 
  \addtoversion{normal}{\mathbfss}{cmss}{bx}{n}
  \addtoversion{bold}{\mathbfss}{cmss}{bx}{n}
  \ifAMStwofonts
    \ifCUPmtlplainloaded \else
      \UseAMStwoboldmath
      \makeatletter
      \new@mathgroup\upmath@group
      \define@mathgroup\mv@normal\upmath@group{eur}{m}{n}
      \define@mathgroup\mv@bold\upmath@group{eur}{b}{n}
      \edef\UPM{\hexnumber\upmath@group}
      \new@mathgroup\amsa@group
      \define@mathgroup\mv@normal\amsa@group{msa}{m}{n}
      \define@mathgroup\mv@bold\amsa@group{msa}{m}{n}
      \edef\AMSa{\hexnumber\amsa@group}
      \makeatother
      \mathchardef\upi="0\UPM19
      \mathchardef\umu="0\UPM16
      \mathchardef\upartial="0\UPM40
      \mathchardef\leqslant="3\AMSa36
      \mathchardef\geqslant="3\AMSa3E

      \let\leq=\leqslant 
      \let\geq=\geqslant \let\ge=\geqslant
    \fi
  \fi
\fi 

\ifnfsstwo
  \DeclareMathAlphabet{\mathbfit}{OT1}{cmr}{bx}{it}
  \SetMathAlphabet\mathbfit{bold}{OT1}{cmr}{bx}{it}
  \DeclareMathAlphabet{\mathbfss}{OT1}{cmss}{bx}{n}
  \SetMathAlphabet\mathbfss{bold}{OT1}{cmss}{bx}{n}
  \ifAMStwofonts
    \ifCUPmtlplainloaded \else
      \DeclareSymbolFont{UPM}{U}{eur}{m}{n}
      \SetSymbolFont{UPM}{bold}{U}{eur}{b}{n}
      \DeclareSymbolFont{AMSa}{U}{msa}{m}{n}
      \DeclareMathSymbol{\upi}{0}{UPM}{"19}
      \DeclareMathSymbol{\umu}{0}{UPM}{"16}
      \DeclareMathSymbol{\upartial}{0}{UPM}{"40}
      \DeclareMathSymbol{\leqslant}{3}{AMSa}{"36}
      \DeclareMathSymbol{\geqslant}{3}{AMSa}{"3E}

      \let\leq=\leqslant 
      \let\geq=\geqslant \let\ge=\geqslant
    \fi
  \fi
\fi 

\ifCUPmtlplainloaded \else
  \ifAMStwofonts \else 
    \def\upi{\pi}
    \def\umu{\mu}
    \def\upartial{\partial}
  \fi
\fi

\def\nd#1#2{{\rm{d} #1 \over \rm{d} #2}}

\title
[Microwave background anisotropies and non-linear structures II]
{Microwave background anisotropies and non-linear structures~~II.
Numerical computations}
\author[Y.~Dabrowski et al.]
{Y.~Dabrowski\thanks{Email: youri@mrao.cam.ac.uk},
M.P.~Hobson, A.N.~Lasenby and C.~Doran\\
Mullard Radio Astronomy Observatory,
Cavendish Laboratory, Madingley Road, Cambridge, CB3 0HE, UK}
\date{Accepted ???. Received ???; in original form \today}
\pagerange{\pageref{firstpage}--\pageref{lastpage}}
\pubyear{1997}

\begin{document}
\newcommand {\modela}{{\it this work (1)}}
\newcommand {\modelb}{{\it this work (2)}}
\newcommand {\modelc}{{\it this work (3)}}
\maketitle
\label{firstpage}

\begin{abstract}
A new method for modelling spherically symmetric inhomogeneities is
applied to the formation of clusters in an expanding Universe.
We impose simple initial velocity and density perturbations
of finite extent and we investigate
the subsequent evolution of the density field.
Photon paths are also calculated,
allowing a detailed consideration of gravitational lensing effects and
microwave background anisotropies induced by the cluster.
We apply the method to modelling high-redshift clusters and,
in particular, we consider the reported microwave decrement
observed towards the quasar pair PC1643+4631 A\&B.
We also consider the effect on the primordial microwave background
power spectrum due to gravitational lensing by a population of massive
high-redshift clusters.
\end{abstract}


\section{Introduction}
\label{intro}

Several methods have been proposed for modelling the formation
of galaxy clusters in an expanding Universe.
Ideally, one would like to model the evolution
of arbitrary matter perturbations in a fully general-relativistic
manner. Unfortunately, this has not proved possible owing to the
complication of the calculations involved, and models
of the formation of clusters of arbitrary shape employ an
approximate linearised approach to incorporate the effects of 
gravity (Mart\'{\i}nez-Gonz\'alez, Sanz \& Silk 1990;
Mart\'{\i}nez-Gonz\'alez \& Sanz 1990; Chodorowski 1991).
Nevertheless, many galaxy clusters can be reasonably
approximated as ellipsoids, and some examples, such as the Coma
cluster, are quasi-spherical. This naturally leads to the use of
spherical symmetry as an approximation, which greatly reduces
the computational complexity and allows the full incorporation of
general-relativistic effects. Early approaches to this problem 
were based on the `Swiss Cheese' model (Rees \& Sciama 1968;
Dyer 1976; Kaiser 1982; Nottale 1982 \& 1984), whereas more recent
attempts have used the continuous Tolman-Bondi solution (Panek 1992;
Arnau et al. 1993; S\'aez, Arnau \& Fullana 1993;
Arnau, Fullana \& S\'aez 1994; Fullana, S\'aez \& Arnau 1994;
Quilis \& S\'aez 1998). 

The Swiss Cheese and Tolman-Bondi approaches both
assume that the cosmological fluid is pressureless. This assumption
may be reasonable for very large structures in which the baryon
content is negligible compared to the dark matter. For typical
clusters, however, the assumption is perhaps questionable
since 10--30 percent of the mass of the cluster may comprise a hot
baryon component, with a non-negligible pressure, that is gravitationally
coupled to the  dark matter.
Nevertheless, calculations by  Quilis, Ib\'a\~nez \& S\'aez (1995),
that include a hot gas component,
show that the effects of pressure are in fact negligible, and that
the collapse of the cluster is well approximated by assuming a
pressureless fluid.
However, the combined use of spherical symmetry
and pressureless assumptions lead inevitably to the unrealistic
situation where the cluster collapses to form a singularity.
Therefore, for the applications described in this paper,
we demand the density profile to be realistic at the time
photons observed today are traversing the cluster.
As pointed out by Quilis et al. (1995),
the assumption of spherical symmetry introduces another difficulty.
As a result of the radial structure of the 
velocity field, gravitational forces generate infalling motions
that are too rapid and lead to fast evolution of the cluster, even
when a hot gas component is included.
An important consequence of this effect is that estimates of
cosmic microwave background (CMB) 
anisotropies due to the cluster are overestimated
and should be regarded as upper limits. However, for high redshift
clusters still in the process of formation, such infalling motions may
be less unrealistic (Navarro, Frenk \& White 1996).

With these points in mind, in this paper we apply a new method for
modelling the evolution of spherically symmetric anisotropies in the
Universe, assuming a pressureless fluid. This method is discussed by
Lasenby et al. (1998) (hereinafter Paper~I) and is based on a new,
gauge-theoretic approach to gravity (Lasenby, Doran \& Gull 1998) that
provides a simple framework in which to investigate spherical cluster
formation. A brief outline of the method is given in Section~\ref{model}
below.
Our intention is to focus on cluster formation at high redshifts.
However, in order to compare our results with existing work we 
first consider low redshift clusters in some detail.
In Sections~\ref{clusterform} and \ref{cmbanisot},
we model the formation of low-redshift clusters with typical
characteristics and calculate the CMB anisotropies expected from these
structures.
Our results are compared with previous studies.
In Section~\ref{highzclusters}, we consider the formation of
high-redshift clusters and discuss their effects on the CMB and
the gravitational lensing of background sources.
As an application, we model a cluster with a view to explaining
the microwave decrement reported towards the quasar pair
PC1643+4631 A\&B (Jones et al. 1997).
Finally, in Section~\ref{cmbpowerspec}, we consider the
effects on the power spectrum of primordial CMB
fluctuations of gravitational lensing by a population of massive
high-redshift clusters.

\section{Theoretical model}
\label{model}

The theoretical model used in this paper is discussed in
detail in Paper~I and provides an exact general-relativistic 
solution describing the evolution of a
spherically-symmetric distribution of pressureless fluid.
The `Newtonian' gauge used for describing this physical system 
is global, having a single time coordinate $t$ and a radial coordinate
$r$. The time coordinate $t$ measures the proper time for observers
comoving with the fluid. In inhomogeneous regions the radial
coordinate $r$ is related to the strength of the tidal force defined by
the Riemann tensor. Throughout this paper we employ natural units
$G=c=\hbar=1$, unless stated otherwise.

The main dynamical variables in this model are the fluid density 
$\rho(t,r)$, its radial velocity field $u(t,r)$ and a generalised
`boost' factor $\Gamma(t,r)$ given by
\begin{equation}
\Gamma^2(t,r) = 1-2M(t,r)/r+u(t,r)^2,
\label{g1def}
\end{equation}
which in the case $M(t,r)=0$ reduces to the square of the standard
special relativistic $\gamma$-factor.
As explained in Paper~I, $M(t,r)$ is the the total 
gravitational mass within a coordinate radius $r$ and is defined
by
\begin{equation}
M(t,r)=\int_0^r 4\pi s^2\rho(t,s) {\rm d}s.
\label{mass}
\end{equation}
Given the density and radial
velocity fields $\rho(t_i,r)$ and $u(t_i,r)$ defined at some initial
time $t_i$, we may calculate $\Gamma(t_i,r)$ and $M(t_i,r)$, which are
then conserved along streamlines. The evolution of the fluid density
and velocity are then easily calculated.

We choose to impose a very simple finite initial perturbation in the
radial velocity
field $u(t_i,r)$ as discussed below.
As explained in Paper~I, assuming that the initial
velocity and density profile arose from primordial perturbations,
we infer the corresponding initial density profile using the relation
\begin{equation}
u(t_i,r)=\frac{2r}{3H(t_i)}\left[2H(t_i)^2-\frac{M(t_i,r)}{r^3}\right],
\label{u_M}
\end{equation}
which, by differentiating~(\ref{mass}), gives
\begin{equation}
\rho(t_i,r)=\frac{3H(t_i)}{8\pi}\left[4H(t_i)-2\frac{u(t_i,r)}{r}-
\frac{\partial u(t_i,r)}{\partial r}\right].
\label{rho_u}
\end{equation}
We note that~(\ref{u_M}) and (\ref{rho_u}) are valid 
for $\Omega(t_i)\approx 1$, which will always be true 
if $t_i$ is soon after inflation.
An important respect in which our approach differs from that of
Panek (1992) or Quilis et al. (1995) is that the
density enhancement resulting from the initial velocity perturbation
is automatically compensated within a finite region by a slightly
under-dense region surrounding it.
Thus the external Universe never feels a gravitational influence from
the perturbed region.
In addition, as explained in Paper~I, imposing an initial velocity
perturbation and working in a fully compensated manner avoid problems
of streamline crossing (i.e. shocks).
\begin{figure}
\centerline{\epsfig{
file=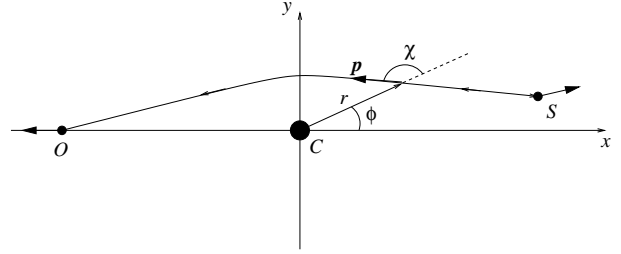,width=8cm}}
\caption{Geometrical arrangement defining the variables used in the
theoretical model. $O$ denotes the observer, $S$ the source
emitting the photon and $C$ the cluster, which is located at the
origin of the coordinate system. Both the observer and the source are
comoving radially with the cosmological fluid. The vector $\bmath{p}$
represents the covariant photon momentum.}
\label{fig2}
\end{figure}

The initial velocity field $u(t_i,r)$ is controlled by four parameters
$R_i$, $a$, $m$ and $H(t_i)$ as follows.
The initial linear extent of the perturbed region is $R_i$ and
the magnitude of the perturbation is controlled by
the velocity gradient at the origin, which we denote by $a$.
The velocity perturbation $u(t_i,r)$ for $0\leq r \leq R_i$ is
defined by a ($2m+1$)-degree polynomial of which the first
$m$ derivatives are matched at the boundaries (i.e. at
$r=0$ and $r=R_i$).
The fourth parameter $H(t_i)$ is
related to the external Universe.
Outside the perturbed region ($r>R_i$), the fluid is described by a
uniformly expanding FRW model.
This is straightforwardly achieved by setting 
$u(t_i,r)=H(t_i)r$, where $H(t_i)$ is the Hubble parameter at the
initial time $t_i$.
As shown in Paper~I, choosing the velocity gradient
at the origin smaller than $H(t_i)$ results in the region $r<R_i$
moving inwards relative to the Hubble flow and eventually collapsing
to form a cluster centred at the origin. 
Conversely, choosing a gradient larger than $H(t_i)$ would lead to
the formation of a void, and the investigation of the evolution of
such structures is also available in this approach. 

Once $R_i$, $a$, $m$ and $H(t_i)$ are fixed, no further parameter
is required to constrain the initial velocity perturbation and the
density perturbation is then fully defined by equation~(\ref{rho_u}).
The initial central density is therefore given by
\begin{equation}
\rho(t_i,0)= \frac{3H(t_i)}{8\pi}\left(4H(t_i)-3a\right),
\label{rho_i}
\end{equation}
and $\rho(t_i,r)$ is represented by a ($2m$)-degree polynomial
for $0<r\leq R_i$.
We note that, at $r=0$, the fluid evolves as a closed FRW Universe with
initial density $\rho (t_i,0)$ and velocity gradient $a$.
Outside the perturbed region, for $r>R_i$, the density is uniform
and equal to
\begin{equation}
\rho_i=\frac{3H^2(t_i)}{8\pi}.
\label{rho_i_b}
\end{equation}
We note that $m$ has to be greater than 1 in order to obtain a sensible
compensated density perturbation with a ($2m$)-degree polynomial.

In addition to the evolution of the fluid, the calculation of
photon trajectories and redshifts is also straightforward.
We can parameterise the photon path in terms of the time parameter
$t$, so that the path is defined by $r(t)$ and $\phi(t)$.
The geometrical configuration adopted is illustrated in Fig.~\ref{fig2}
where $\chi$ is the angle between the photon's position vector and its
covariant momentum $\bmath{p}$. 
As explained in Paper~I, $\chi$ is an observable quantity and is also
equal to the angle between
the photon trajectory and the cluster centre, as measured by observers
comoving with the fluid.
Given an initial position $(r_0,\phi_0)$ and direction $\chi_0$,
the set of first-order differential equations given in Paper~I 
may be integrated numerically to obtain the subsequent photon path
and frequency.

\section{Cluster formation}
\label{clusterform}

\subsection{Initial conditions}
\label{initcond}

Using the model outlined above, it is straightforward to investigate
the formation of spherical clusters.
We begin specifying the initial
conditions such that the resulting cluster is similar to those
observed. First, we must choose the time $t_i$ at which the
initial perturbation is applied. 
Throughout this paper we take $t_i$ to represent the epoch $z=10^3$
and we assume that the Universe is at critical density so that
$\Omega(t)=1$ for all values of $t$. Unless stated otherwise, we perform
simulations for $H(t_0) = 100~{\rm km~s^{-1}~Mpc^{-1}}$ where $t_0$ is
the present time.
Results will be quoted using the reduced Hubble parameter $h(t)$ which
is equal to $H(t)$ expressed in units of $100~{\rm km~s^{-1}~Mpc^{-1}}$.
Furthermore, quantities evaluated at $t=t_0$ will be written simply
with a zero subscript, e.g. $\Omega(t_0)=\Omega_0$, $h(t_0)=h_0$, and
so on. The free parameters of our model are $H(t_i)$, which
defines the external Universe, and $a$, $R_i$ and $m$, which determine the
initial perturbation. The following paragraphs (\ref{init1},
\ref{init2} \& \ref{init3}) describe how we fix these parameters.
\begin{figure}
\centerline{\epsfig{
file=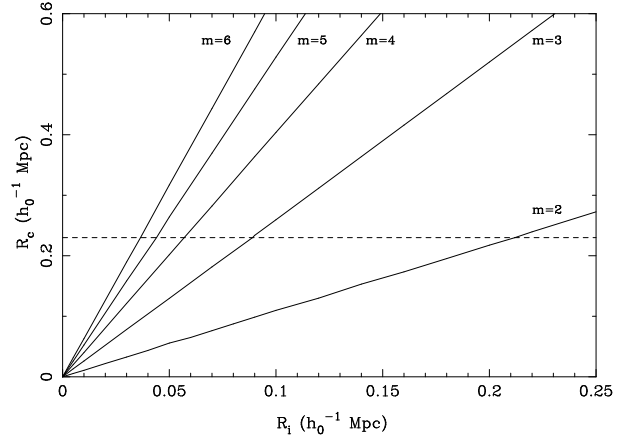,angle=-90,width=8cm}}
\caption{The cluster core radius $R_{\rm c}$ at $z=0.09$ as a function
of the initial perturbation size $R_i$ for various values of the integer
parameter $m$ (the latter controls the polynomial order - see text).
The points of intersection between the dashed line and the solid
lines indicate the suitable value of $R_i$ for each case.}
\label{fig_rc_ri}
\end{figure}
\begin{figure}
\centerline{\epsfig{
file=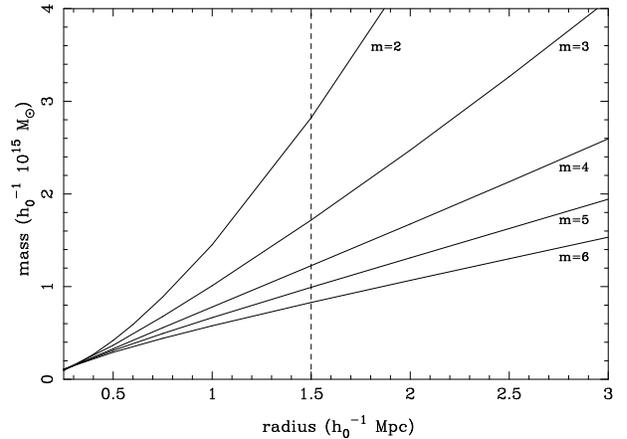,angle=-90,width=8cm}}
\caption{The total gravitational cluster mass at $z=0.09$ within a sphere
of radius $r$ in unit of $10^{15}~h_0^{-1}~{\rm M_{\odot}}$,
as a function of $r$ for various values of
$m$. $R_i$ has been fixed using results from Figure~\ref{fig_rc_ri}.}
\label{fig_mass_r}
\end{figure}
\subsubsection {External Universe}
\label{init1}
Given $\Omega_0=1$ and $h_0$, the parameter $H(t_i)$ is
fully defined. For $t_i$ representing $z=10^3$, we find 
$H(t_i)=3.167\times 10^6 h_0~{\rm km~s^{-1}~Mpc^{-1}}$.

\subsubsection{Final appearance of the cluster}
\label{init2}
The only guide to fixing the initial parameters $a$, $R_i$ and $m$
is the final properties of the cluster, as observed today.
In this section, we model a very rich Abell cluster,
in order to compare our results with those obtained by Quilis et al.
(1995) and Panek (1992). As our standard configuration, we consider
a cluster at a redshift $z=0.09$, with a core radius
$R_{\rm c} = 0.23 h_0^{-1}~{\rm Mpc}$ (where $R_{\rm c}$ is defined as the
radius at which the cluster density falls to one-half its
maximum value). A redshift $z=0.09$ corresponds to
an effective distance of $D=250 h^{-1}_0~{\rm Mpc}$.
The central baryonic number density of the cluster is
taken to be $n_{\rm c}=10^4 h_0^{1/2}~{\rm m^{-3}}$. We further
assume that, at all times, the baryon component contributes
$10 h_0^{-3/2}$ per cent of the total mass, the remainder being dark
matter.
Finally, because we are interested in rich clusters, the total
gravitational mass $M_{\rm c}$   within $r<R_{\rm c}$ is taken to be
greater than $10^{15} h^{-1}_0~{\rm M_{\odot}}$.
Note that the values of $R_{\rm c}$, $n_{\rm c}$ and $M_{\rm c}$
are defined at the epoch when a photon observed today was
traversing the cluster.

We should discuss our choices regarding the $h$-dependences
stated above.
The $h$-dependence of the core radius ensures that the
observed angular size of the cluster is independent of the Hubble
parameter and equal here to 3.4 arcmin.
The $h$-dependence of the central baryonic number density is chosen
so that the cluster's X-ray flux is also independent of $h_0$
(Peebles 1993; Jones \& Forman 1984).
Hence the three {\it observational} properties of the cluster,
namely its redshift, angular size and X-ray flux do not depend
on our choice of $h_0$. The $h$-dependence of the dark matter
ratio gives a total gravitational mass which scales as $h_0^{-1}$
as expected within the isothermal sphere approximation (e.g. Peebles
1993). In this case the total density follows the power law
form $\rho\propto 1/r^2$ (see Section~\ref{charac}) and
the baryon mass alone scales as $h_0^{-5/2}$.

We note that the distance $D=250 h^{-1}_0~{\rm Mpc}$ is
somewhat larger than the value of $100 h^{-1}_0~{\rm Mpc}$ used by
Quilis et al. (1995).
This is because our simulations show that in the case
of the cluster described above, but located at a distance of
$100 h^{-1}_0~{\rm Mpc}$, the observer is marginally decoupled from
the Hubble flow (particularly for the $m=2$ model).
Thus in order that the observer resides in the external Universe
it was necessary to place the cluster at a larger distance
(see Section~\ref{cmbanisot}).

\subsubsection {Fixing the initial perturbation}
\label{init3}

\begin{figure}
\centerline{\epsfig{
file=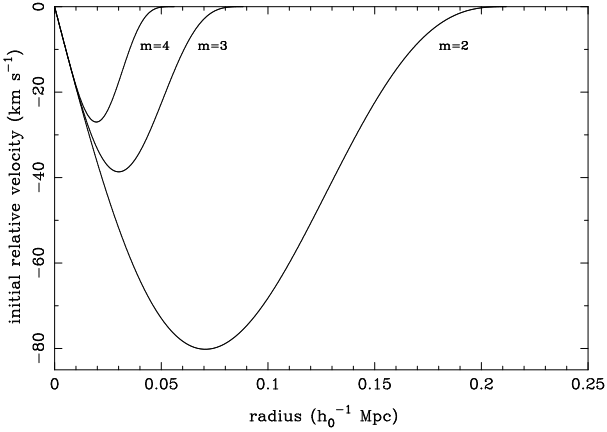,angle=-90,width=8cm}}
\caption{Initial relative velocity $u(t_i,r)-H(t_i)r$ for
the $m=2,3$ and $4$ models at $z=10^3$.}
\label{fig_init_vel}
\end{figure}

\begin{figure}
\centerline{\epsfig{
file=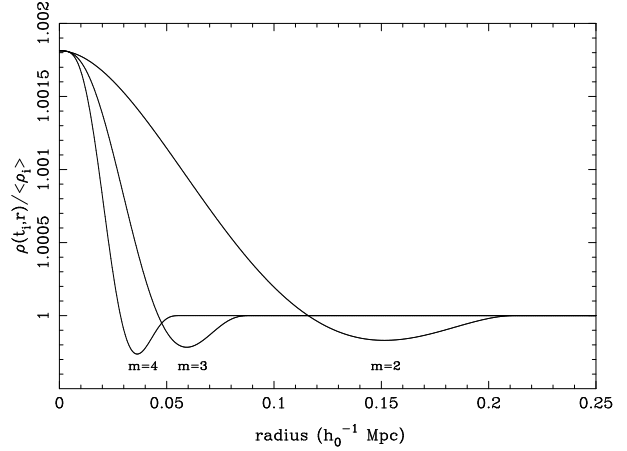,angle=-90,width=8cm}}
\caption{Initial density profiles $\rho(t_i,r)$ for
the $m=2,3$ and $4$ models at $z=10^3$.}
\label{fig_init_rho}
\end{figure}

At the origin ($r=0$), the fluid evolves like a closed
FRW Universe with initial density $\rho (t_i,0)$ given in (\ref{rho_i})
and velocity gradient $a$. Therefore $a$ is directly constrained by
requiring that $n_{\rm c}=10^4 h_0^{1/2}~{\rm m^{-3}}$ at $z=0.09$.
We find that $1-\left[a/H(t_i)\right]\sim 6\times 10^{-4}$.
The further requirements, namely $R_{\rm c} = 0.23 h_0^{-1}~{\rm Mpc}$
and $M_{\rm c}\geq  10^{15} h^{-1}_0~{\rm M_{\odot}}$ at $z=0.09$,
that allow us to fix both $R_i$ and $m$ as follows.
We find (see Fig.~\ref{fig_rc_ri}) that in order to satisfy
$R_{\rm c} = 0.23 h_0^{-1}~{\rm Mpc}$, $R_i$ should be equal to
$0.21 h_0^{-1}$, $0.089 h_0^{-1}$, $0.057 h_0^{-1}$, $0.044 h_0^{-1}$
and $0.036 h_0^{-1}~{\rm Mpc}$ for $m=2$, $3$, $4$, $5$ and $6$,
respectively.
Because we require that $M_{\rm c}\geq  10^{15} h_0^{-1}~{\rm M_{\odot}}$
at $z=0.09$, we can see from Fig.~\ref{fig_mass_r} that only
$m=2$, $m=3$ and $m=4$ cases are suitable.
The initial velocity and density perturbations for the $m=2,3$ and $4$
models are shown in Figs.~\ref{fig_init_vel} and \ref{fig_init_rho}
respectively.
Each of these perturbation give rise to the same cluster
properties at $z=0.09$, namely:
$n_{\rm c}=10^4 h_0^{1/2}~{\rm m^{-3}}$,
$R_{\rm c} = 0.23 h_0^{-1}~{\rm Mpc}$ and
$M_{\rm c}\geq  10^{15} h_0^{-1}~{\rm M_{\odot}}$.
We note that the $m=2$ model requires a much larger perturbation.
This could be explained by the fact that the initial
density profile for larger $m$ is flatter at the origin,
concentrating more mass at the centre and therefore
allowing for a faster collapse.
As a typical model, most of the figures presented in this paper are for
$m=3$ while all the numerical results of Sections~\ref{fluidevol},
\ref{charac} and \ref{cmbanisot} will be quoted for the three cases
$m=2,3$ and $4$.

\subsection{Fluid evolution}
\label{fluidevol}
\begin{figure}
\centerline{\epsfig{
file=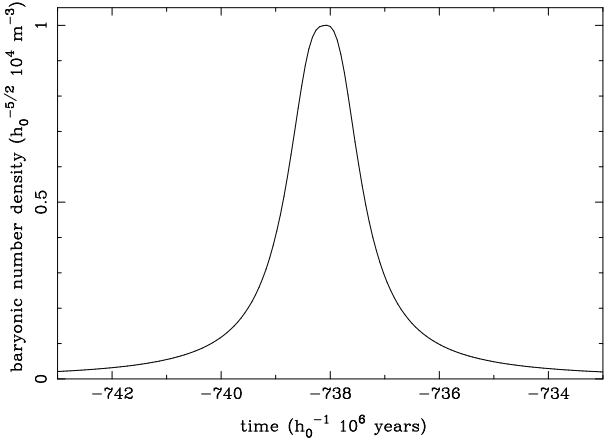,angle=-90,
width=8cm}}
\caption{The baryonic number density, as a function of cosmic time
$t$, along a photon path traversing the centre of the cluster.
The time $t=0$ corresponds to the current epoch and we assume $m=3$.}
\label{fig3}
\end{figure}

\begin{figure}
\centerline{\epsfig{
file=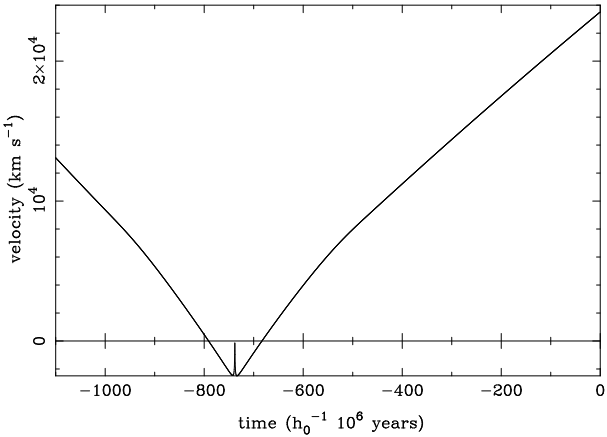,angle=-90,
width=8cm}}
\caption{The radial velocity field $u(t,r)$ along the photon trajectory.
The time $t=0$ corresponds to the current epoch and we assume $m=3$.}
\label{fig4}
\end{figure}

\begin{figure}
\centerline{\epsfig{
file=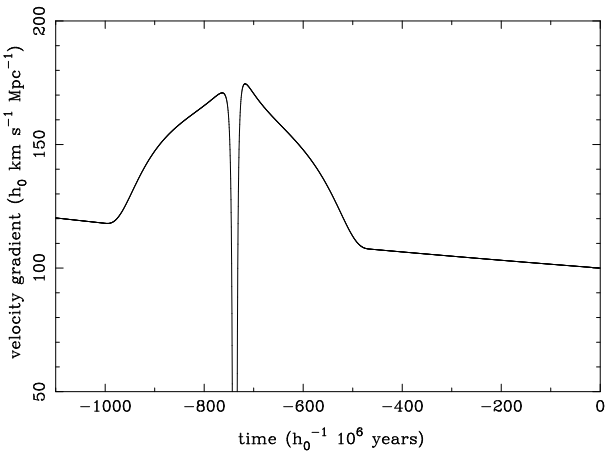,angle=-90,
width=8cm}}
\caption{As in Fig.~\ref{fig4}, but for the velocity gradient
$\partial u/\partial r$ along the photon path.}
\label{fig5}
\end{figure}

\begin{figure}
\centerline{\epsfig{
file=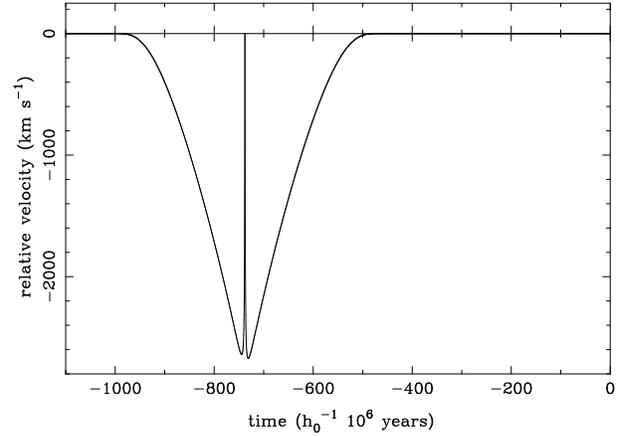,angle=-90,
width=8cm}}
\caption{As in Fig.~\ref{fig4}, but for the relative velocity 
$u(t,r)-H(t)r$ along the photon path.}
\label{fig5b}
\end{figure}

In order to appreciate the results presented in this
paper, in particular the effect on the CMB induced by the evolving
cluster, we present a series of figures (Figs.~\ref{fig3},
\ref{fig4}, \ref{fig5} and \ref{fig5b}) illustrating various fluid
quantities as measured by a set of observers comoving with the fluid,
on the path of a photon traversing the centre of the cluster.
These figures allow a check that our model is behaving sensibly.

Fig.~\ref{fig3} shows the baryonic number density, as a function of cosmic
time $t$, along the path of a photon which travels straight through the
centre of the cluster and reaches the observer at the current epoch $t_0$;
the curve displayed is for $m=3$.
For convenience we take the current epoch $t_0=0$, so
that all times are expressed in Myr prior to today.  From the figure
we see that the photon traversed the cluster in approximately 5~Myr,
encountering a maximum baryonic number density of
$10^4h_0^{-5/2}~{\rm m^{-3}}$
at a time $t_c=738 h_0^{-1}~{\rm Myr}$ ago.

Similarly, Figs.~\ref{fig4} \& \ref{fig5} show 
respectively the fluid radial velocity field $u(t,r)$ and velocity
gradient $\partial u/\partial r$ along the photon path for $m=3$.
Several features in these plots should be noted. It is clear from
Fig.~\ref{fig5} that from about $500 h_0^{-1}$~Myr prior to the current
epoch, the velocity gradient is merely that due to
the slowing expansion of the external Universe, and so the photon is
free-streaming towards the observer.  As required, this velocity
gradient tends to the value $H_0 = 100 h_0~{\rm km~s^{-1}~Mpc^{-1}}$ at
$t=0$. From Fig.~\ref{fig4}, we see that, even though the photon
experiences a perturbation due to the cluster up to $500 h_0^{-1}$~Myr
prior to today, it is only when in the inner part of the cluster,
between $810 h_0^{-1}$ and $675 h_0^{-1}$~Myr ago, that the fluid
velocity  is directed inwards.
In the outer parts of the perturbation the fluid is still moving outwards,
although it is, of course, collapsing with respect to the Hubble flow.
The fluid velocity relative to the Hubble flow is shown in
Fig.~\ref{fig5b}.

For models with $m\neq 3$, the corresponding results are very similar
to those shown in Figs.~\ref{fig3}, \ref{fig4}, \ref{fig5} and
\ref{fig5b}, but with a
larger or smaller perturbed region for the models $m=2$ and $m=4$
respectively.

\subsection{Cluster characteristics}
\label{charac}

Before giving numerical values of the cluster characteristics and
comparing them with previous work, we first check that the obtained
cluster profile has a realistic shape. We consider here a broad
family of density profiles (Zhao 1996, Kravtsov et al. 1998) 
given by
\begin{equation}
n(r)=\frac{n_0}{\left(\frac{r}{R}\right)^{\gamma}
\left[1+\left(\frac{r}{R}\right)^{\alpha}\right]^{
\frac{3\beta-\gamma}{\alpha}}}.
\label{profile}
\end{equation}
We note that this family includes the three most widely used types
of density distribution:
(i) Profiles derived from $N$-body codes and proposed by
Navarro et al. (1996),
which diverge at the origin, hereafter NFW profiles.
These correspond to $(\alpha,\beta,\gamma)=(1,1,1)$ in equation
(\ref{profile}). For this case, $n_0/4$ is the density at
$r=R$;
(ii) Isothermal profiles, also known as King profiles (Rood et al. 1972,
Sarazin 1988), which have a finite central density $n=n_0$
and for which $(\alpha,\beta,\gamma)=(2,2/3,0)$.
Here, the density profile follows the very simple form
\begin{equation}
n(r)=\frac{n_0}{1+\left(\frac{r}{R}\right)^2},
\label{profileking}
\end{equation}
and $R$ is the radius at which $n=n_0/2$;
(iii) We define here $\beta$-models as being a generalisation of King
profiles. They are widely used to fit observed profiles and rotation
curves and correspond to $(\alpha,\beta,\gamma)=(2,\beta,0)$.

\begin{figure}
\centerline{\epsfig{
file=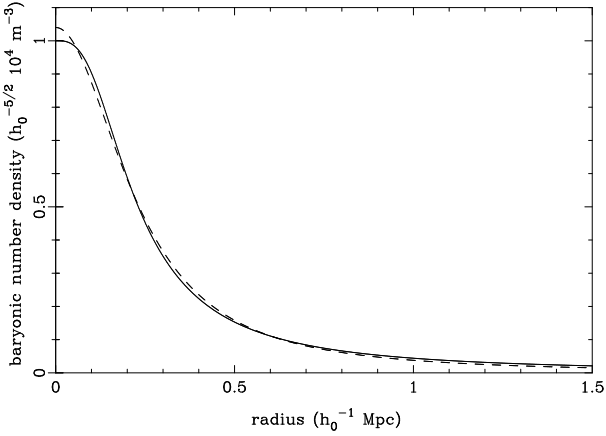,angle=-90,
width=8cm}}
\caption{The baryonic number density of the cluster (solid line) as a
function of radius at the time $t_c=738 h_0^{-1}$~Myr ago, when a photon
traversing the centre of the cluster experiences the maximum density;
$m=3$ is assumed. The dashed line is the best-fit $\beta$-model
(see text).}
\label{fig6}
\end{figure}

\begin{figure}
\centerline{\epsfig{
file=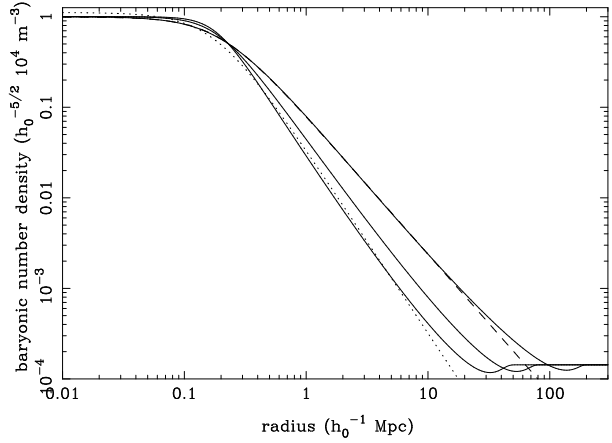,angle=-90,
width=8cm}}
\caption{
Log-log plot of the baryonic number density of the clusters (solid lines)
as a function of radius at the time $t_c=738 h_0^{-1}$~Myr ago.
Simulations are for $m=2,3$ and $4$ from right to left.
The dashed line is the best-fit $\beta$-model for the $m=2$ case,
while the dotted line is for the $m=4$ case (see text).
}
\label{fig_dens10}
\end{figure}

In order to compare our simulated cluster profiles with 
equation (\ref{profile}),
we consider the variation with radius $r$ of the number density
$n_{\rm c}(r)$ of our clusters at the time
$t_{\rm c}=738 h_0^{-1}~{\rm Myr}$ ago,
when the photon reaches the point of maximum density along its trajectory.
This profile is plotted as the solid line in Fig.~\ref{fig6} 
for the case $m=3$.
Performing a simple least-squares fit for the three parameters of the
$\beta$-models from $r=0$ to $r=1.5 h_0^{-1}$~Mpc, we find
$n_0 =1.04\pm 0.08\times h_0^{-5/2} 10^4~{\rm m^{-3}}$, 
$R = 0.25\pm 0.04 h_0^{-1}~{\rm Mpc}$ and 
$\beta=0.75\pm 0.15$, where the quoted errors are one-sigma limits.
The best-fit $\beta$-model is shown as the dashed line in Fig.~\ref{fig6}.
The $m=4$ model gives rise to a density profile somewhat flatter at
the origin than in the $m=3$ case of Fig.~\ref{fig6}, resulting in a
slightly less accurate fit to the $\beta$-model. However, for the
$m=2$ case, the obtained simulated profile is almost indistinguishable
from its $\beta$-model best-fit from $r=0$ up to $r\sim 10 h_0^{-1}$~Mpc.
This can be seen in the log-log plot of Fig.~\ref{fig_dens10},
where the density
profiles of models $m=2,3$ and $4$ are plotted in solid lines for
$0.01 h_0^{-1}~{\rm Mpc}\leq r\leq 300 h_0^{-1}~{\rm Mpc}$.
The dashed line in Fig.~\ref{fig_dens10} corresponds to the best-fit
$\beta$-model for $m=2$. In this case, we find
$n_0 =0.99\pm 0.15\times h_0^{-5/2} 10^4~{\rm m^{-3}}$, 
$R = 0.19\pm 0.03 h_0^{-1}~{\rm Mpc}$ and 
$\beta=0.5\pm 0.05$. The best-fit $\beta$-model for the $m=4$ case
is the dotted line of Fig.~\ref{fig_dens10} and corresponds to
$n_0 =1.11\pm 0.14\times h_0^{-5/2} 10^4~{\rm m^{-3}}$, 
$R = 0.18\pm 0.03 h_0^{-1}~{\rm Mpc}$ and 
$\beta=0.67\pm 0.05$ which is equivalent to the King profile described
in equation~(\ref{profileking}).

\begin{figure}
\centerline{\epsfig{
file=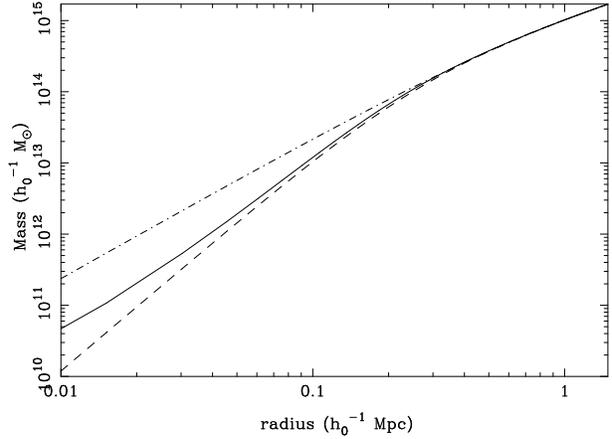,angle=-90,
width=8cm}}
\caption{The total gravitational mass of the cluster included within
a given radius (solid line) at the the time $t_c=738 h_0^{-1}$~Myr ago,
when a photon traversing the centre of the cluster experiences the
maximum density; $m=3$ is assumed. The dashed line is the best-fit
King profile while the dot-dashed line is the best-fit NFW profile.
}
\label{fig6b}
\end{figure}

We note that the central density of our simulated cluster
is well defined, unlike
some of the $(\alpha,\beta,\gamma)$ profiles of equation (\ref{profile}),
which diverge at $r=0$.
Therefore, in addition to the density distribution, we choose to fit the 
mass profile $M(r)$ defined by equation (\ref{mass}) at the time
$t=t_{\rm c}$. For both the analytical profiles
defined by equation (\ref{profile}) and our simulated cluster, the
mass density is defined by $\rho(r)=10 h_0^{-3/2} m_{\rm p}\times n(r)$, 
where $m_{\rm p}$ is the 
proton mass and the factor $10 h_0^{-3/2}$ is the ratio of the total
mass to the baryonic mass.
The mass profile obtained from our model, together with the best-fit
King and NFW profiles from $r=0$ to $r=1.5 h_0^{-1}$~Mpc,
is shown in Fig.~\ref{fig6b} for the case $m=3$.
We note that our model agrees with masses derived from both
King and NFW profiles for $r>0.4 h_0^{-1}~{\rm Mpc}$ while a King model is
favoured for small radii. This is not surprising since both the King
model and our simulated density profile
are finite at the origin, whereas the NFW profile diverges at $r=0$.
A least-squares fit of the all the parameters of equation (\ref{profile})
gives $(\alpha,\beta,\gamma)=(1.6,0.67,0.3)$ which seems to favour
the simple King profile of equation~(\ref{profileking}). In this case
the number density approaches the power law $n\propto 1/r^2$ which is
in agreement with the isothermal sphere approximation
that we made in Section~\ref{init2} regarding the $h$-dependences.


It is a particularly satisfying feature of our approach that the
radial density profile of the perturbation evolves into one that
matches quite closely those of observed clusters and clusters produced
in $N$-body simulations.
It should be emphasised again that the initial conditions assumed a
very simple finite velocity perturbation with an associated density
perturbation.
It is very encouraging that these straightforward initial conditions
can lead to such realistic density profiles as that presented
in Figs.~\ref{fig6}, \ref{fig_dens10} and \ref{fig6b}.

As mentioned above, we define the distance to the cluster as $D=250
h_0^{-1}~{\rm Mpc}$ and the cluster core radius to be
$R_c = 0.23h_0^{-1}~{\rm Mpc}$.
From our simulations, however, we may also measure other
important parameters describing the cluster at the time when
the photon passes through its centre.
Of particular interest is the turnaround
radius $R_t$, which is defined as the radius at which
the fluid velocity changes from being radially inwards to radially
outwards. Another important characteristic is the ratio $\delta$ of
the cluster central density to that of the external Universe.  We find
$R_t=34.2 h_0^{-1}$, $R_t=16.5 h_0^{-1}$ and $R_t=11.4 h_0^{-1}$~Mpc
for $m=2,3$ and $4$ respectively.
The density ratio found in all three cases is
$\delta=7\times 10^3$.

We may compare these basic properties with those found by
Panek (1992) and Quilis et al. (1995). 
In making such a comparison, however, we must
remember that these authors specify different initial conditions.
The reader is referred to their respective papers for details of
the assumed initial density profiles.
In Table~\ref{table1}, we compare the properties
of the resulting clusters for Panek type I and type II models and
the Quilis et al. model.
\begin{table}
\caption{Previous estimates of cluster properties. $D$ denotes the 
distance to
the cluster, $R_c$ is the cluster core radius, $R_t$ is the turnaround
radius and $\delta$ is the ratio of the central cluster density to
that of the external Universe. All distances are quoted in
$h_0^{-1}~{\rm Mpc}$}
\begin{center}
\begin{tabular}{lcccc} \hline
Author              & $D$ & $R_c$ & $R_t$ & $\delta$ \\ \hline
Panek I             & 100 & 0.30  & 11.7  & 674 \\
Panek II            & 100 & 1.00  & 18.8  & 674 \\
Quilis et al.        & 100 & 0.23  & ---   & 3220\\ 
This work ($m=2$) & 250 & 0.23  & 34.2  & 7000\\ 
This work ($m=3$) & 250 & 0.23  & 16.5  & 7000\\ 
This work ($m=4$) & 250 & 0.23  & 11.4  & 7000\\ 
\hline
\end{tabular}
\end{center}
\label{table1}
\end{table}

We see from the table that the turnaround radii for the previous model
clusters are in reasonable agreement with our estimate for
$h_0=1$. However, the density ratio $\delta$ derived from our
simulations in this case is over twice that for the Quilis et al.
model and over a factor of ten greater than those quoted for the two Panek
models. This suggests that a velocity perturbation of the form
assumed here is in fact more effective in
producing highly non-linear structures than the initial density
perturbations assumed by previous authors.

We may also calculate the total mass of the cluster contained within
spheres of various radii and compare the result with previous
calculations. These are given in Table~\ref{table2}.
\begin{table}
\caption{The total gravitational mass contained within spheres of
given radii for a spherical cluster. $M_{m=2,3,4}(r)$ gives the mass
calculated in this work for $m=2,3$ and $4$ models,
whereas $M_{\rm PI}(r)$, $M_{\rm PII}(r)$ and 
$M_{\rm Q}(r)$ refer respectively to estimates for the two Panek models
and the Quilis et al. model.
All masses are quoted in $10^{15} h_0^{-1}~{\rm M_{\odot}}$ and radii
are in $h_0^{-1}$~Mpc.}
\begin{center}
\begin{tabular}{ccccc} \hline
$r$              & 1    & 1.5  & 2    & 4    \\ \hline
$M_{m=2}(r)$     & 1.45 & 2.81 & 4.42 & 12.6 \\
$M_{m=3}(r)$     & 1.01 & 1.72 & 2.47 & 5.84 \\
$M_{m=4}(r)$     & 0.78 & 1.22 & 1.67 & 3.55 \\
$M_{\rm PI}(r)$  & --   &  --  &  --  & 1.8  \\   
$M_{\rm PII}(r)$ & --   &  --  &  --  & 5.7  \\
$M_{\rm Q}(r)$   & --   & 0.95 &  2.0 & 6.2  \\ \hline
\end{tabular}
\end{center}
\label{table2}
\end{table}
The derived masses for models $m=2,3$ and $4$ and agree reasonably
well with Quilis et al. and Panek, particularly for $m=3$.

In concluding this section, it must be noted that at later times than
those illustrated above, the subsequent evolution of the cluster is
rather unrealistic.
As a result of the assumptions of a pressureless fluid and a radial
velocity distribution, the collapse of the cluster continues unchecked,
leading to the ultimate formation of a singularity. Hence the
formation of stable clusters is not admitted by this model. Nevertheless,
there are no difficulties in studying photon propagation beyond the
time the singularity forms, as long as the photon has passed the origin
by this point.
Indeed, it is very useful to be able to consider
photons that passed through the cluster shortly before the singularity
formed, but which were not received by the observer until some time later.
The inclusion of a baryonic component with pressure will be discussed
in a forthcoming paper.

\section{CMB anisotropy}
\label{cmbanisot}

Using the photon propagation equations (Paper~I),
we calculate the predicted CMB anisotropy produced
when a photon passes through the collapsing cluster. As discussed
above, however, the assumption of a pressureless fluid and spherical
symmetry result in the cluster evolution being too rapid. Therefore
the anisotropies derived in this section should properly be considered
as upper limits.

For our present purposes let us ignore primordial fluctuations in the
CMB and concentrate on secondary anisotropies caused by the interaction
of CMB photons with non-linear structures such as clusters. 
Three main secondary anisotropies can occur: (i)
the kinetic Sunyaev-Zel'dovich (SZ) effect resulting from the
peculiar bulk velocity of a cluster relative to the observer, which
results in the Doppler shift of the CMB blackbody spectrum; (ii)
the thermal SZ effect produced by the inverse
Compton scattering of CMB photons to higher energies by hot electrons
in the cluster gas; (iii) a non-linear gravitational effect on the CMB
photon as it passes through the cluster. In this
section, we will be concerned with the last of these, which
is often called the Rees-Sciama effect (Rees \& Sciama 1968).

 A simple, approximate way to understand why such an effect should
occur is to consider the gravitational potential well experienced by
the CMB photon. Clearly, if the cluster is collapsing as the photon
passes through it, then the photon must `climb out' of a well deeper
than that into which it `fell', resulting in a net redshift of the
photon. However, a compensating effect also occurs, since a photon
passing through the potential well is delayed with respect to a photon
that did not pass through the cluster. Thus, if these two photons arrive
simultaneously at the observer, the one that passed through the
cluster must have been emitted at an earlier time, when the Universe
was hotter, leading to a blueshift of the photon. Thus there is an
interplay between these two competing effects, and the overall
effect can be of either sign depending on the details of the evolution
of the potential well. However, as seen in Paper~I (equations 33 and 36),
the gauge theory approach offers a clear alternative understanding of the
physical processes involved in the Rees-Sciama effect, showing that the
microwave decrement is directly related to the velocity gradient
$\partial u(t,r)/\partial r$ encountered by the
photon rather than the potential well variation.
This raises questions on issues related to the generalisation
of available models to the non-spherical case and models including
a pressure component as discussed in Quilis \& S\'aez (1998).

\begin{figure}
\centerline{\epsfig{
file=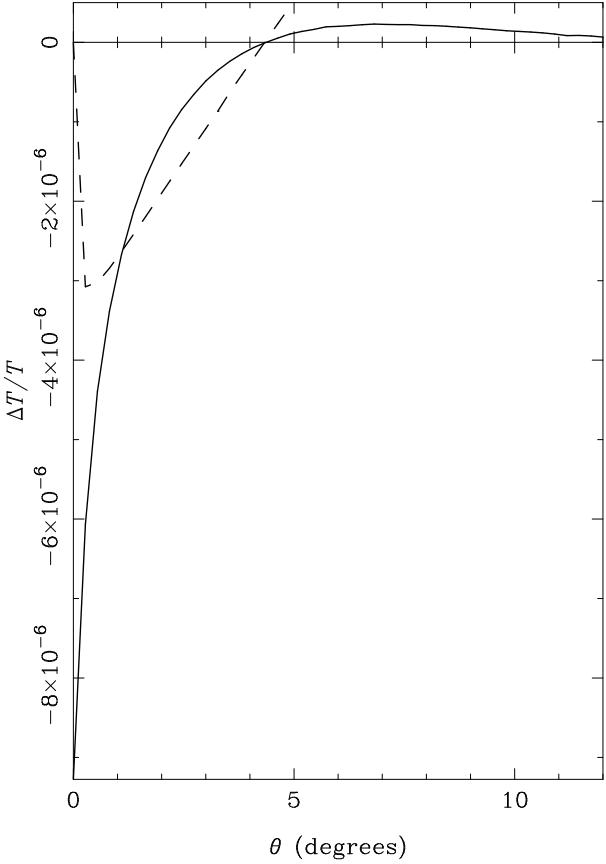,
width=8cm}}
\caption{The predicted CMB anisotropy (solid line) due to the
gravitational effect of the cluster
described in Section~\ref{clusterform}, for $m=3$.
The anisotropy is plotted in units of $\Delta T/T$ as a function of
the projected angle $\theta$ on the sky from the centre of the
cluster.
The dashed line shows the fluid velocity in arbitrary units 
at the point of maximum density along any given photon trajectory.}
\label{fig7}
\end{figure}

Fig.~\ref{fig7} shows the gravitational CMB anisotropy 
due to the cluster described in the last section, for $m=3$.
The anisotropy is plotted in $\Delta T/T$ as a function of projected
angle $\theta$ on the sky from the centre of the cluster. 
\begin{figure}
\centerline{\epsfig{
file=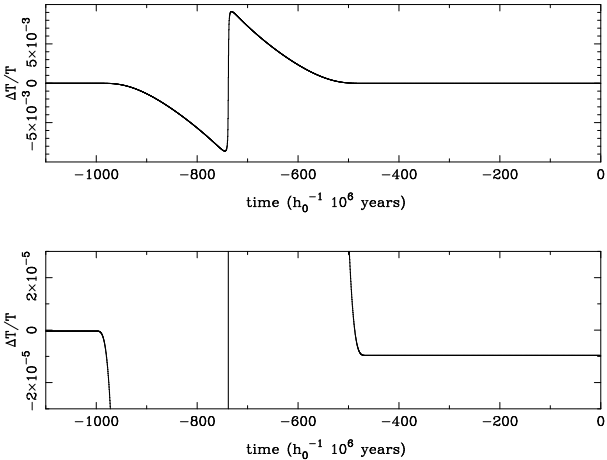,
angle=-90,width=8cm}}
\caption{Top: the change in $\Delta T/T$ for a CMB photon passing through
the centre of the cluster, as seen by an observer comoving with the
fluid at any given time for the $m=3$ model.
Bottom: the same curve plotted between
smaller limits. $\Delta T$ for observers inside the cluster is defined
with respect to the temperature of the external Universe at the same time.
The time $t=0$ corresponds to the current epoch.}
\label{fig7b}
\end{figure}
The maximum anisotropy occurs at the centre of the cluster and has the
value $\Delta T/T = -0.93\times 10^{-5}$.
The decrement decreases rapidly with projected angle on the sky and
reaches zero at $\theta=4.4$ degrees.
For larger angles the anisotropy becomes slightly positive
indicating a net blueshift of the CMB photons. This positive effect
reaches a maximum of $\Delta T/T = 0.25\times 10^{-6}$ at $\theta=6.8$
degrees before slowly tending towards zero as $\theta$ increases.
We note that, using the $h$-dependences stated in Section~\ref{init2},
the obtained $\Delta T/T$ is independent of $h_0$.
Also plotted in Fig.~\ref{fig7} is the velocity in arbitrary units
at the point of maximum density
along any given photon trajectory. It is interesting to note that the
crossover between the negative CMB anisotropy and the positive one at
$\theta=5$ degrees occurs close to the point at which the velocity 
changes from being radially inwards to radially outwards. 

Anisotropies of a similar shape are also found for the $m=2$ and $4$
cases and we find the maximum decrements to be
$\Delta T/T = -5.8\times 10^{-5}$ and $\Delta T/T = -0.4\times 10^{-5}$
respectively. The result for $m=2$ is somewhat larger
than that found for $m=3$, which is to be expected since the cluster
obtained for $m=2$ is larger even though it satisfies the same
observational requirements stated in Section~\ref{init2}.

\begin{table}
\caption{Various estimates of the anisotropy $\Delta T/T$
compared with our three estimations for models $m=2,3$ and $4$.}
\begin{center}
\begin{tabular}{lr} \hline
Authors&
$\Delta T/T$ \\ \hline
Panek type I       & $-0.15\times 10^{-5}$ \\
Panek type II      & $-0.6\times 10^{-5}$ \\
Quilis et al.      & $-1.2\times 10^{-5}$ \\
Chodorowski I      & $-0.26\times 10^{-5}$ \\
Chodorowski II     & $-0.77\times 10^{-5}$ \\
Nottale            & $\approx -10\times  10^{-5}$ \\
This work ($m=2$)  & $-5.8\times 10^{-5}$ \\
This work ($m=3$)  & $-0.93\times 10^{-5}$ \\ 
This work ($m=4$)  & $-0.4\times 10^{-5}$ \\ \hline
\end{tabular}
\end{center}
\label{tabledt}
\end{table}

We can compare the central decrements calculated above with those of
previous authors. Results are reported in Table~\ref{tabledt}
for Panek (1992) type I and type II clusters, Quilis et al. (1995),
Chodorowski (1991) models I and II and finally for 
Nottale (1984), who used the Swiss Cheese model to predict a
considerably larger central decrement of
$\Delta T/T \approx -10\times 10^{-5}$.
However, this last value corresponds to very dense,
rapidly collapsing objects and the resulting cluster mass is
much greater than those observed.

We see that our predicted central decrement for $m=3$
is in rough agreement with that quoted by Quilis et al., which is
slightly larger than those found by Panek and Chodorowski.
However, as discussed above, the effect predicted here has to be
regarded as an upper limit, particularly in our $m=2$ model
where $\Delta T/T = -5.8\times 10^{-5}$.

In Fig.~\ref{fig7b}, we plot the change in $\Delta T/T$ of a CMB photon
passing through the centre of the cluster, as seen by an observer comoving
with the fluid at any given time, for the $m=3$ model.
The temperature difference is measured 
relative to CMB photons that have not interacted with a cluster.
The top plot illustrates the wide
range of anisotropy produced by the interaction with the cluster. The
bottom plot displays the same curve plotted between smaller limits and
shows the resulting $\Delta T/T = -0.93\times 10^{-5}$ as discussed above.
We note that $\Delta T/T\sim 0$ at the origin which corresponds to
$t_{\rm c}=738 h_0^{-1}~{\rm Myr}$ ago.
This plot also illustrates that $\Delta T/T$ does not change when the
photon is in the region exterior to the cluster.
Therefore, as long as the observer lies in the external Universe
(i.e. sufficiently far away from the cluster centre), the magnitude
of the Rees-Sciama effect is independent of the distance
to the cluster. In the case where the observer lies sufficiently close
to the cluster that he is decoupled from the Hubble flow, predictions
of the Rees-Sciama effect magnitude could be highly inaccurate, as seen
from the large gradients of Fig.~\ref{fig7b} (bottom panel). In the
case where $m=2$, the sharp gradient in $\Delta T/T$ that occurs at
$t\sim500h_0^{-1}$~Myr ago in Fig.~\ref{fig7b} is shifted to a time
close to the present day.
Indeed, this is the reason why we had to place our cluster at an effective
distance of $D=250h_0^{-1}$~Mpc (see Section~\ref{init2}).

\section{Application to high-redshift clusters}
\label{highzclusters}

We have so far concentrated on the formation of rich, low-redshift 
Abell clusters in order to compare our results with previous authors.
We have commented, however, that such clusters are most likely
virialised to some extent and therefore are not well modelled
by a spherical free-fall collapse. Nevertheless, such a model may
provide a better description of the dynamical state of high-redshift
clusters. We would expect clusters at $z \ga 1$ still to be in the
process of formation, and therefore may be reasonably approximated 
as quasi-spherical objects with large radial infall velocities.
Massive clusters at high redshift have recently been identified,
either by X-ray measurements, gravitational lensing or spectroscopic
identification ($z=0.83$, Luppino and Kaiser 1997; $z=0.996$, 
Deltorn et al. 1997).

The existence and distribution of high redshift clusters are of
great importance regarding the discrimination between different
cosmological scenarios.
The observational aspect of the problem has developed only recently.
Indeed identifying bound structures at high
redshift is a real observational challenge not only because of the
decrease in brightness with distance but also because of the high degree
of confusion due to the projection of numerous foreground and background
objects. As a search strategy, Jones et al. (1997) proposed
to use the Sunyaev-Zel'dovich (SZ) effect to detect the presence
of high-redshift clusters, the main argument being that the 
magnitude of the SZ effect is independent of redshift.
Since the Rees-Sciama and the SZ effects usually add up to form
a total observed CMB decrement, it is essential to know
precisely the magnitude of the Rees-Sciama effect so that the
SZ contribution can be evaluated accurately.
Therefore, a model such as ours, which allows us to
understand the details of the Rees-Sciama effect, should prove useful
in the context of programmes such as that carried out by
Jones et al. (1997) or similar future work.

In order to apply our model to some on-going observations,
we choose to investigate the properties of high-redshift
clusters which may model
the microwave decrement reported
towards the quasar pair PC1643+4631 A\&B (Jones et al. 1997).
This decrement is observed at 15 GHz and lies between a pair of quasars
at redshifts $z=3.79$ and $z=3.83$ separated by 198~arcsec on the sky.
Jones et al. assume the decrement to be a thermal Sunyaev-Zel'dovich
effect due to an intervening cluster and estimate the minimum central
temperature decrement to be $\Delta T/T = -2.1\times 10^{-4}$.
However, no cluster is evident in either
ROSAT X-ray observations or infrared R-, J- and K-band images taken by
the WHT and the UKIRT (Saunders et al. 1997). The magnitude limits
obtained lead the authors to suggest that any cluster causing the
decrement must lie at $z \ga 1$. 

Modelling of the SZ observations, assuming $h_0=0.5$, suggests that
the putative cluster 
has a core radius of $R_c=0.3-0.5~{\rm Mpc}$ and a gas
mass inside a 2-Mpc radius of $2.0-3.5\times 10^{14}~{\rm M_{\odot}}$.
This implies a total mass, including both luminous and dark
matter, greater than $10^{15}~{\rm M_{\odot}}$.
Saunders et al. further point out
that for such a massive cluster at $z=1-2$, the Einstein ring radius
for sources at $z\approx 3.8$ is approximately 100~arcsec. Indeed,
simple gravitational lens modelling can simultaneously produce the 
observed SZ effect and make the true positions of the two quasars
virtually coincident. The suggestion that the observed quasar pair are
in fact images of the same object is supported by the remarkable
similarity between the quasar spectra, apart from their one-percent 
redshift difference. Saunders et al. suggest that this difference 
could be explained in terms
of quasar evolution over a delay between the two lightpaths
of $\sim 10^3~{\rm yrs}$.
A further similar observation has been reported by Richards et al. (1997)
where a microwave decrement is also detected towards a pair of quasars.

Using the model described in Section~\ref{model}, we now consider
the formation of a massive cluster located at $z=1$ and
investigate both the gravitational lensing effects and the microwave
decrement produced by such a cluster

\subsection{Gravitational lensing}
\label{sec_lens}
\begin{figure}
\centerline{\epsfig{
file=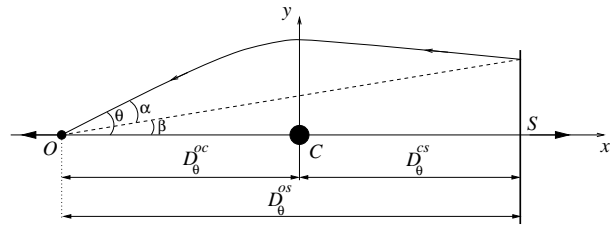,
width=8cm}}
\caption{
Geometrical arrangement of a source-lens system. $S$ denotes 
the plane of the source, which is comoving with the Hubble flow.
A photon is emitted from an arbitrary position on this plane
towards the observer $O$. The emission point is defined by the
observed angle $\theta$ (in presence of the lens) or by the observed
angle $\beta$ (in absence of the lens) and $\alpha=\theta-\beta$.
$D_\theta^{xy}$ denotes the angular-size distance from the point $x$
to the point $y$.
}
\label{lens_geom}
\end{figure}
We calculate the evolution of the mass distribution and the photon
paths as described in Sections~\ref{model} and \ref{clusterform} using
the following parameters: $\Omega(t)=1$;
$H(t_0)=50~{\rm km~s^{-1}~Mpc^{-1}}$;
the maximum baryonic number density along the 
photon path is taken to be $10^4~{\rm m^{-3}}$; the total
gravitational mass over baryonic mass ratio is fixed to $10$;
the initial perturbation is for $m=3$;
the observed cluster redshift is taken to be $z=1$ and the core radius is
0.45~Mpc (52~arcsec).
The source of photons corresponding to the observed
pair of quasars is comoving with the Hubble flow and lies at a
redshift $z=3.8$.
We note that the assumptions made here concerning the size and density
of the cluster are somewhat different than in Section~\ref{init2}.
In order to fit PC1643+4631~A\&B observations, we find it necessary to
model such a rich cluster.

\begin{figure}
\centerline{\epsfig{
file=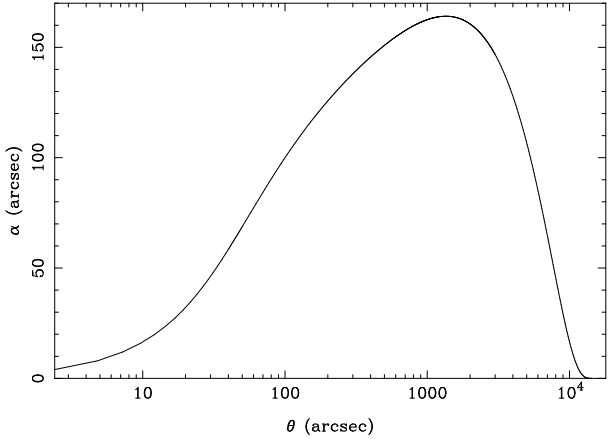,angle=-90,
width=8cm}}
\caption{The reduced deflection angle $\alpha$ as a function of viewing
angle $\theta$. The cluster lies at $z=1$ and the source at $z=3.8$.
}
\label{fig8}
\end{figure}

We now consider the simple case where a (slightly extended) source
lies with its centre on the same line
of sight as the cluster. For a
spherically symmetric cluster, the resulting image
is radially symmetric and any position in the image is defined by
the radial angle $\theta$ measured from the cluster centre.
We can define the {\it reduced} deflection angle
$\alpha(\theta)$ by using the standard
{\it lens equation} (Refsdal, 1964; Blandford \& Narayan, 1992): 
\begin{equation}
\alpha(\theta)= \theta-\beta(\theta),
\label{lens}
\end{equation}
where $\theta$ and $\beta$ are the observed viewing angles of the
same point on the source plane with and without the lensing mass
respectively.
We note that in the context of our model described in Section~\ref{model},
$\theta$ and $\beta$ are the relevant values of the angle $\chi$, as
observed by the comoving observer $O$ of Fig.~\ref{lens_geom}.
In the geometrical arrangement of Fig.~\ref{lens_geom}, the {\it reduced}
deflection angle can be read as the {\it convergence} (i.e. radial
stretching) of the source.
For various image positions $\theta$, we compute
the corresponding angle $\beta$ and so the convergence $\alpha(\theta)$
in the case of our evolving gravitational lens.
We find that the lensing effect occurs up to rather large angles
( $\sim 3$ degrees), as seen in Fig.~\ref{fig8}.

\begin{figure}
\centerline{\epsfig{
file=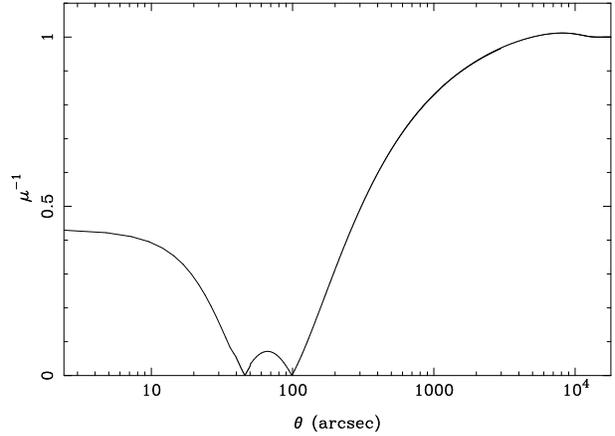,angle=-90,
width=8cm}}
\caption{
The inverse of the magnification factor as a function of the viewing
angle $\theta$.
The magnification becomes infinite at $\theta_c \simeq 46''$ and
at $\theta_e \simeq 100''$.
}
\label{fig9}
\end{figure}

Another interesting characteristic of our source-lens system is its
magnification factor $\mu(\theta)$ defined to be the ratio between the
observed solid angles of the same unit area at the source plane, with and
without the lens. Using (\ref{lens}) in the circularly symmetric lens
case, we find
\begin{equation}
\frac{1}{\mu(\theta)}=[\theta-\alpha(\theta)]
\frac{\left[1-\nd{\alpha(\theta)}{\theta}\right]}{\theta}.
\label{magnify}
\end{equation}
Using Fig.~\ref{fig8} we may compute the corresponding magnification,
which is displayed in Fig.~\ref{fig9}.
We can see from (\ref{magnify}) that there are two cases
where the magnification diverges. First, on the Einstein ring radius
$\theta_e$, we have $\theta_e-\alpha(\theta_e)=\beta(\theta_e)=0$.
Any photon observed with $\theta=\theta_e$ originates from the
same emission point at the centre of the source plane. Therefore
the solid angle observed in the absence of the lens shrinks to zero
and the magnification becomes infinite.
Second, the magnification also diverges
when ${\rm d}\alpha/{\rm d}\theta =1$ (in this case we have
${\rm d}\beta/{\rm d}\theta=0$); we denote the angle at which this
occurs by $\theta_c$. 
\begin{figure}
\centerline{\epsfig{
file=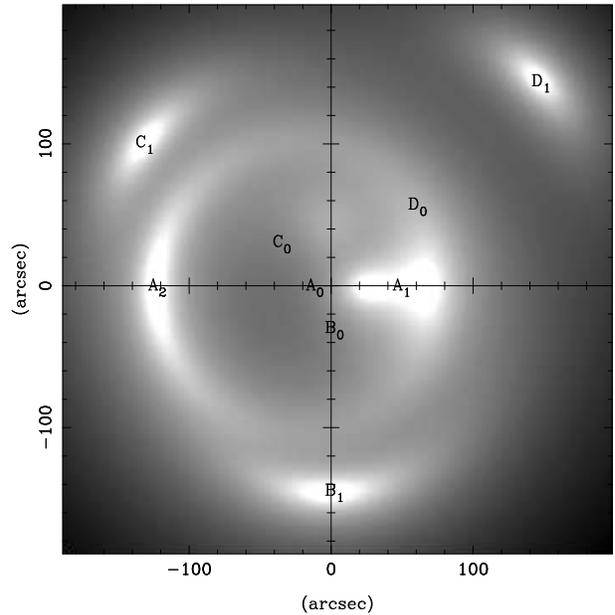,angle=-90,
width=8cm}}
\caption{
Lensing image of four sources A, B, C \& D located at the same
redshift $z=3.8$. ${\rm A_0}$, ${\rm B_0}$, ${\rm C_0}$, ${\rm D_0}$
denotes the positions of the sources as observed in the absence of the
cluster. ${\rm A_1}$, ${\rm A_2}$, ${\rm B_1}$, ${\rm C_1}$ and
${\rm D_1}$ are the corresponding
lensed images.
}
\label{fig10}
\end{figure}

On the source plane, the caustic defined by the first case is a
critical point whereas the caustic defined by the second is a
ring. These caustics are called the {\it point caustic} and {\it outer
caustic}, respectively. A point source at the point caustic produces a
ring of radius $\theta_e$ on the image plane (i.e. the Einstein ring);
a point source on the outer caustic is observed as a double source with
one radially elongated and reversed image at $\theta=\theta_c$ and a
second tangentially elongated image at $\theta > \theta_e$. In general
a point source located outside the outer caustic produces a unique
lensed image while a triple image is obtained if the source lies
within the outer caustic (Blandford \& Narayan, 1992).

As an illustration of the lensing properties of our cluster, in
Fig.~\ref{fig10} we show a lensed image of four extended sources
A, B, C \& D. Each source is located at the same redshift
$z=3.8$ and is assumed to be circularly symmetric.  Sources
B, C, D lie outside the outer caustic whereas A is positioned on
the caustic itself.  We note the double image (almost triple on the
figure) produced by A and its tangential/radial characteristics. The
positions of the sources as observed without the lens are indicated so
that the lensing effect of the cluster can be fully appreciated.

It is also interesting to note the effect of the lens on a 
background of sources at a given redshift with a projected
number density (in the absence of the lens) of $N_0(\theta)$.
In the presence of the lens the observed number density becomes
$N(\theta)=N_0(\theta)/\mu$, and from Fig.~\ref{fig9} we see that
$N(\theta) < N_0(\theta)$ for $\theta \la 1$ degree.

\subsection{Microwave decrement}
\label{sec_cmb}

\begin{figure}
\centerline{\epsfig{
file=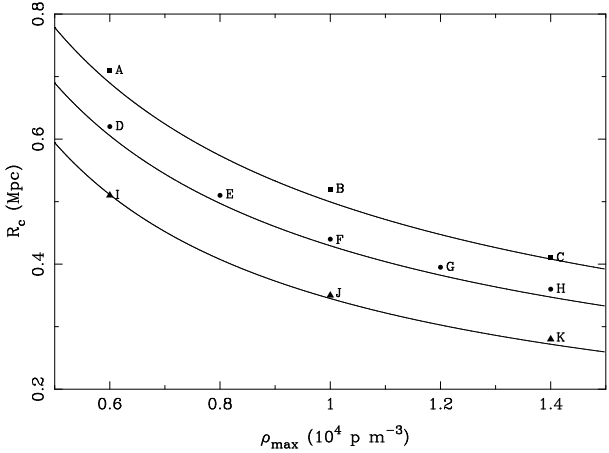,angle=-90,
width=8cm}}
\caption{
Einstein ring radius contours over a range of $\rho_{\rm max}$ and $R_c$
values for, from top to bottom, $\theta_e=150''$, $\theta_e=100''$ and
$\theta_e=50''$. The solid lines denotes the contours computed assuming a
simple static lens with an King density profile. The points are exact 
calculations for an evolving lens 
(squares for $\theta_e=150''$, circles for $\theta_e=100''$ and
triangles for $\theta_e=50''$).
}
\label{fig12}
\end{figure}
\begin{table}
\caption{Properties of clusters in Fig.~\ref{fig12}. $\theta_e$ denotes
the Einstein ring radius; $T_e$ is the electron temperature in the
cluster; $\Delta T_{\rm SZ}$ is the Sunyaev-Zel'dovich decrement and
$\Delta T_{\rm RS}$ is the Rees-Sciama decrement.}
\begin{center}
\begin{tabular}{ccccc} \hline
Cluster & $\theta_e$ & $T_e~{\rm (K)}$ &
$\Delta T_{\rm SZ}$ (${\rm \mu K}$) & $\Delta T_{\rm RS}$
(${\rm \mu K}$)\\ 
\hline 
A & $150''$ & $2.12\times 10^7$ & 523 & 306 \\
B & $150''$ & $1.54\times 10^7$ & 466 & 283 \\
C & $150''$ & $1.36\times 10^7$ & 459 & 274 \\
\hline
D & $100''$ & $2.27\times 10^7$ & 491 & 201 \\
E & $100''$ & $1.95\times 10^7$ & 464 & 185 \\
F & $100''$ & $1.78\times 10^7$ & 458 & 178 \\
G & $100''$ & $1.63\times 10^7$ & 459 & 172 \\
H & $100''$ & $1.56\times 10^7$ & 463 & 169 \\
\hline
I & $50''$ & $2.64\times 10^7$ & 464 & 117 \\
J & $50''$ & $2.26\times 10^7$ & 466 & 87 \\
K & $50''$ & $2.12\times 10^7$ & 490 & 79 \\
\hline
\end{tabular}
\end{center}
\label{table3}
\end{table}

In this section we investigate the contribution of the Rees-Sciama 
effect to the total temperature decrement reported towards the quasar
pair PC1643+4631 A\&B and discuss the possibility that the quasars
are gravitationally lensed images of the same object (Dabrowski et al.
1997; Dabrowski 1997).

In order to obtain any quantitative results we need to constrain our model
parameters. We suppose that both quasars have a redshift $z=3.8$ and
that the cluster lies at $z=1$. As for Section~\ref{sec_lens},
we assume that $\Omega(t)=1$,
$H(t_0)=50~{\rm km~s^{-1}~Mpc^{-1}}$ and that the total gravitational
mass over baryonic mass ratio is $10$.
The initial perturbation is for $m=3$.
Two parameters remains to be fixed: $\rho_{\rm max}$, the maximum baryonic
density encountered by an observed photon and the characteristic
core radius $R_c$ of the cluster at the time the photon reaches the
point of maximum density.
To fix these parameters, we have two observational constraints:
the quasar pair separation of 198~arcsec and the total flux observed by
the Ryle Telescope of $-380~{\rm \mu Jy}$ (Jones et al. 1997).

We first consider the lensing separation. Naturally, our
spherically-symmetric model cannot account for multiple images that do
not lie on the same radial line from the centre of the cluster, as in
the case of PC1643+4631~A\&B.
We note here that a more suitable elliptical setup might present
a longer projected line of sight and so enhance the CMB decrement effect.
For the purpose of this paper, we consider the
Einstein diameter $2\times\theta_e$ to be characteristic of the lensing
power of the cluster and compare this with the separation of the 
quasar images.
For each pair of values $\rho_{\rm max}$ and $R_c$, which describe the
cluster central density and core radius, we obtain a value for the
Einstein ring radius $\theta_e$. The solid lines
in Fig.~\ref{fig12} are
contours in $(\rho_{\rm max},R_c)$-space corresponding to
Einstein ring radii of $\theta_e=50$, 100 and 150~arcsec respectively,
calculated without using our model but
assuming a static lens with a density distribution described by
a King profile (see equation~\ref{profile}).
In this case the Einstein ring radius may be found simply by
solving numerically the standard equation
\begin{equation}
\theta_e=\sqrt{\frac{4GM(\theta_e)}{c^2}\frac{D_\theta^{oc}}
{D_\theta^{os}D_\theta^{cs}}},
\label{einstein}
\end{equation}
where $M(\theta_e)$ is the projected mass included within the radius
$\theta_e$, $D_\theta^{oc}$ denotes the angular-size distance from the
observer to the cluster, $D_\theta^{os}$ the angular-size distance
from the observer to the source and $D_\theta^{cs}$ the angular-size
distance from the cluster to the source (see Fig.~\ref{lens_geom}).
These distances were computed using the formulae given by
Blandford \& Narayan (1992).

In Fig.~\ref{fig12}, we also plot 11 selected cluster setups that
have been computed using exact calculations from our evolving spherical
cluster model. Since these points represent evolving lenses,
and the corresponding clusters do not have exact King density
profiles, we see that these points lie slightly away from the contours.
The points plotted have been chosen so that A, B, C denote
lenses for which $\theta_e=150''$; D, E, F, G, H have 
$\theta_e=100''$ and I, J, K have $\theta_e=50''$.

Secondly, we consider the total flux observed by the Ryle telescope
between the pair of quasars.
Table~\ref{table3} shows the electron temperatures $T_e$ required in
order that the clusters in Fig.~\ref{fig12} each
produce a total observed SZ flux of $-380~{\rm \mu Jy}$.
Also listed in the table are the corresponding SZ and Rees-Sciama 
temperature decrements for each cluster.
We note that the ratio of the Rees-Sciama effect
as compared to the SZ effect varies from $0.61$ for cluster B to
$0.16$ for cluster K. Thus in most cases the Rees-Sciama effect is not
negligible and might contribute significantly to the total observed CMB
decrement. As pointed out in the Introduction, however, 
we note that these values are to be taken as upper limits.
\begin{figure*}
\centerline{\epsfig{
file=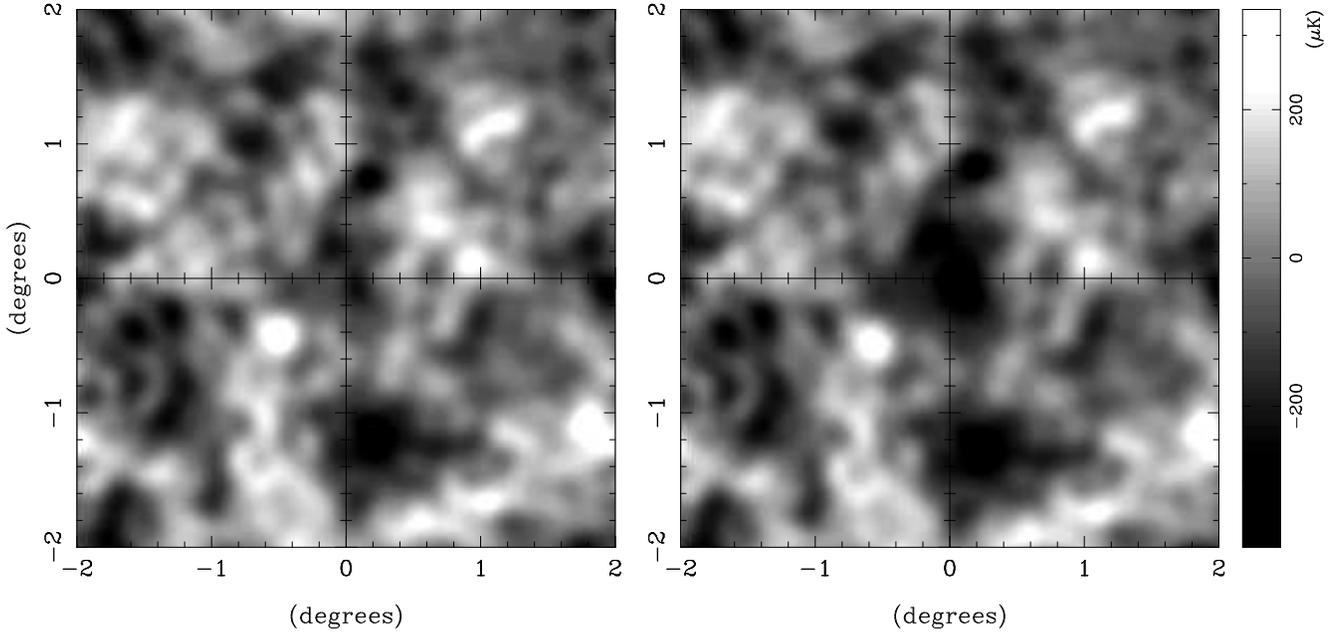,
width=17.6cm}}
\caption{Left: a realisation of CMB fluctuations in a standard
inflationary CDM model. Right: the same realisation but with a rich
cluster at its centre, producing a Rees-Sciama temperature decrement
and gravitational lensing effects.}
\label{figcmb}
\end{figure*}

We take cluster F as a typical
example of a cluster which may explain the observations of Jones et al.
(1997). The total mass contained within spheres of various radii
for cluster F are: $7.8\times 10^{15}~{\rm M_\odot}$ within 2~Mpc;
$1.9\times 10^{16}~{\rm M_\odot}$ within 4~Mpc;
and $4.6\times 10^{16}~{\rm M_\odot}$ within 8~Mpc. The ratio of the
central density to the density of the external Universe at the time
the photon passed through the cluster is $\delta = 4440$ and the
turnaround radius $R_t = 25.3~{\rm Mpc}$.

We may also consider the possible time delay between the light paths
for the two quasar images PC1643+4631 A\&B. A reasonable approximation
may be obtained in our model by considering a source located directly
behind the cluster at $z=3.8$ and calculating the time delay between a
photon from the source travelling straight through the centre of the
cluster and one which follows a lensed path, appearing at the Einstein
ring radius $\theta_e$.  We note that our model takes full account of
all the relativistic effects caused by the motions of the source and
observer in the Hubble flow and the gravitational effects of the
collapsing cluster. We find that the proper time delay in the frame of
the source is $\approx 150$ years. This is rather a brief period for
the quasar to evolve sufficiently to explain the small redshift
difference of the two quasar images.

Finally, we note that observations of the PC1643+4631  field are still
in progress. For example, Haynes et al. (1998) carried out
deep optical imaging with the William Herschel Telescope
and show that the excess of faint blue galaxies in the
field might be consistent with a cluster lying at $z\sim 2$.
Furthermore, very recent ROSAT observations may suggest that 
any intervening cluster producing the temperature decrement towards
the quasar pair PC1643+4631 A\&B should in fact have a redshift $z > 3$
(Kneissl 1997; Kneissl, Sunyaev and White 1998).
We have therefore repeated our analysis for a cluster
lying at a redshift greater than 1 and chosen the (maybe conservative)
value of $z=2$.
In this case we consider a cluster with a baryonic number density
of $10^4~{\rm m^{-3}}$ and a core radius $R_c = 0.74~{\rm Mpc}$;
this ensures an Einstein ring radius of 100~arcsec.
The total mass contained within a sphere of 2~Mpc is
$1.74\times 10^{16}~{\rm M_\odot}$, the density
ratio $\delta = 1313$ and the turnaround radius $R_t = 22~{\rm Mpc}$.
In order to retrieve the observed SZ flux of $-380~{\rm \mu Jy}$,
we require the electron gas
temperature of the cluster to be $1.3\times 10^7~{\rm K}$. This gives
an SZ temperature decrement of $547~{\rm \mu K}$, while the upper
limit of the Rees-Sciama temperature decrement is in this
case $\approx 600~{\rm \mu K}$
(i.e. it is of the same order as the thermal SZ effect).
For such a cluster, we find that the lensing time delay, as described in
the previous paragraph, is approximately 230 years in the frame of
the emitting source placed at a redshift $z=3.8$.

\section{Gravitational lensing of primordial CMB anisotropies}
\label{cmbpowerspec}

We discussed in the previous section 
a possible unified model for the formation of a high-redshift cluster
with the properties required to produce a microwave decrement and
gravitational lensing effects consistent with the observations 
towards the quasar pair PC1643+4631 A\&B. It is also of some interest,
however, to investigate the effects of such a cluster on primordial CMB
fluctuations.

In this Section, we therefore consider
the non-linear gravitational effects on a typical spectrum
of primordial fluctuations of the massive cluster F, discussed in the
last section, at a redshift $z=1$. These
effects include both the gravitational lensing and shift in energy of
CMB photons passing through the cluster. We do not, however, include 
any SZ effect that the cluster may produce.

The effect on primordial fluctuations is investigated by propagating
through the evolving cluster a population of CMB photons due to a
typical primordial CMB field. 
It will be seen from Fig.~\ref{fig8} that the reduced deflection angle
produced by this
cluster falls to zero only at large angles of about 3 degrees. 
We therefore take as our typical primordial CMB field a
$4\times4$-degree realisation of CMB fluctuations in a standard
inflationary Universe dominated by cold dark matter.

Fig.~\ref{figcmb} shows that effect of the cluster lying at the centre
of the field of primordial CMB fluctuations. We see that the cluster
produces a pronounced Rees-Sciama decrement as well as causing a slight
radial stretching of the CMB fluctuations which extends as far as the
edge of the field. 

\subsection{Massive cluster abundances}
\label{abundances}

We may also consider the effect of a population of such clusters
on the CMB power spectrum.
As seen in previous work (e.g. Eke, Cole \& Frenk 1996) standard models
using the Press-Schechter formalism in a $\Omega_0=1$ Universe
predict no structure of mass $\sim 10^{15}M_{\odot}$ at
redshift greater than $z\sim 0.8$. However counts of clusters are
sensitive to $\Omega_0$ so that open cosmological models with
$\Omega_0 \sim 0.3$ allow the presence of massive clusters at $z>1$.
Eke et al. (1996)
extended the Press-Schechter formalism to flat cosmological models with
a cosmological constant $\Lambda$ so that $\Omega_{\Lambda}+\Omega_0=1$,
where $\Omega_{\Lambda}=\Lambda/(3H_0^2)$.
For a given value of $\Omega_0$, the redshift distribution of massive
clusters is similar in both open and flat cosmological models.
We assume here $\Omega_{\Lambda}+\Omega_0=1$ and a cold dark matter model.
We have used the formalism and normalisation described
in Eke et al. (1996). The number density of clusters with mass
between $M$ and $M+dM$ at a given redshift $z$ is given by $n(z,M)dM$
with 
\begin{equation}
n(z,M)=\left(\frac{2}{\pi}\right)^{\frac{1}{2}}
\left(\frac{M}{M_8}\right)^{\alpha}\frac{\bar{\rho_0}\alpha\delta_c(z)}
{\sigma_8 M^2}
\exp^{-\frac{1}{2}\left[\frac{\delta_c(z)}{\sigma_8}\left(\frac{M}{M_8}
\right)^{\alpha}\right]^2} 
\label{press-sch}
\end{equation}
where $M_8$ is the mass contained in spheres of radius $8h_0^{-1}$~Mpc;
$\sigma_8=0.52\Omega_0^{-0.52+0.13\Omega_0}$ is the rms linear
fluctuation amplitude within $8h_0^{-1}$~Mpc spheres;
\begin{equation}
\bar{\rho_0}=\frac{3H_0^2}{8\pi{\rm G}}\left(1-\Omega_{\Lambda}\right)
\end{equation}
is the present mean density of the Universe;
$\alpha=(0.68+0.4\Gamma)/3$ (see White, Efstathiou and Frenk 1993) for
$\Gamma=\Omega_0 h_0=0.25$; and $\delta_c(z)$ is the density threshold
for collapse as derived in Eke et al. (1996) in the case
$\Omega_{\Lambda}+\Omega_0=1$.
The number of clusters of mass from $M$ to $M+dM$ within a
spherical shell extending from $z$ to $z+dz$ is given by $n(z,M)dMdV$
with $dV=4\pi D^2dr$ where $D$ is the effective distance and $dr$ is
the comoving radial coordinate element equal to $-(1+z)cdt$. 
Therefore
\begin{equation}
dV=-4\pi cD^2(1+z)\frac{\partial{t}}{\partial{z}}dz
\label{dV}
\end{equation}
where $\partial{t}/\partial{z}$ and $D$ are derived from
Fukugita et al. (1992) for the $\Omega_{\Lambda}+\Omega_0=1$ case:
\begin{equation}
\frac{\partial{t}}{\partial{z}}=-
\left[H_0(1+z)\sqrt{3\Omega_0z(1+z)+1+\Omega_0z^3}\right]^{-1},
\label{dtdz}
\end{equation}
and
\begin{equation}
D=\frac{c}{H_0}\int_0^z{\frac{dz}{\sqrt{\Omega_0(1+z)^3+(1-\Omega_0)}}}
\label{D}
\end{equation}

\begin{figure}
\centerline{\epsfig{
file=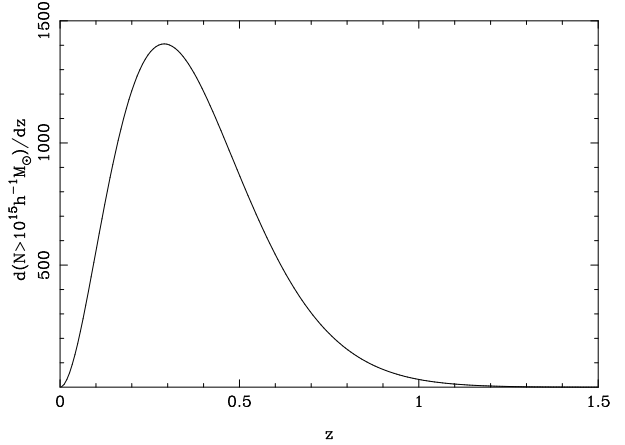,angle=-90,
width=8cm}}
\caption{
The number of clusters with mass larger than $10^{15}h_0^{-1}M_{\odot}$
per unit redshift in the whole sky, as predicted by the Press-Schechter
formalism.
We assume $\Omega_0=0.3$, $\Omega_0+\Omega_{\Lambda}=1$ and the
normalisation of Eke et al. (1996)
}
\label{fig_press1}
\end{figure}

\begin{figure}
\centerline{\epsfig{
file=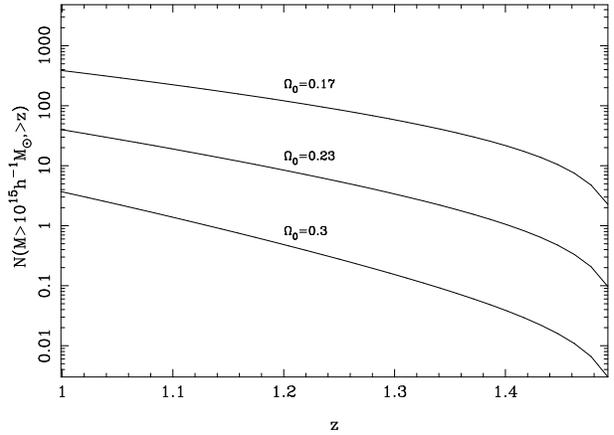,angle=-90,
width=8cm}}
\caption{
The number of clusters with mass larger than $10^{15}h_0^{-1}M_{\odot}$
at a distance further than the given redshift $z$ in the whole sky,
as predicted by the Press-Schechter formalism.
The lower, middle and upper curves are for $\Omega_0=0.3$,
$\Omega_0=0.23$ and $\Omega_0=0.17$ respectively and
$\Omega_0+\Omega_{\Lambda}=1$ is assumed in each cases.
The normalisation is that of Eke et al. (1996)
}
\label{fig_press2}
\end{figure}

We use the above formulae and find the distributions given
in Figs.~\ref{fig_press1} \&  \ref{fig_press2}.
Fig.~\ref{fig_press1} shows that for $\Omega_0=0.3$ only 
very few clusters more massive than $10^{15}M_{\odot}$ are expected at
redshift larger than 1.
Fig.~\ref{fig_press2} gives the estimated number of $10^{15}M_{\odot}$
clusters in the whole sky as a function of redshift, for various value
of $\Omega_0$.
For $\Omega_0=0.3$, $0.23$ and $0.17$ there are, in the whole sky,
$\sim 4$, $40$ and $400$ clusters respectively at $z>1$ and more massive
than $10^{15}M_{\odot}$.
We note that the presence of massive objects as distant as PC1643
may suggest that results using Press-Schechter formalism in a standard
cosmology underestimate massive cluster counts at high redshift.
This is the reason why, in addition to
the realistic assumption $\Omega_0=0.3$ (e.g. Webster et al. 1998), we
are considering here $\Omega_0=0.23$ and $\Omega_0=0.17$ which give
abundances respectively $10$ times and $100$ times higher than those
expected for $\Omega_0=0.3$. 

\subsection{Effect on the CMB power spectrum}

\begin{figure}
\centerline{\epsfig{
file=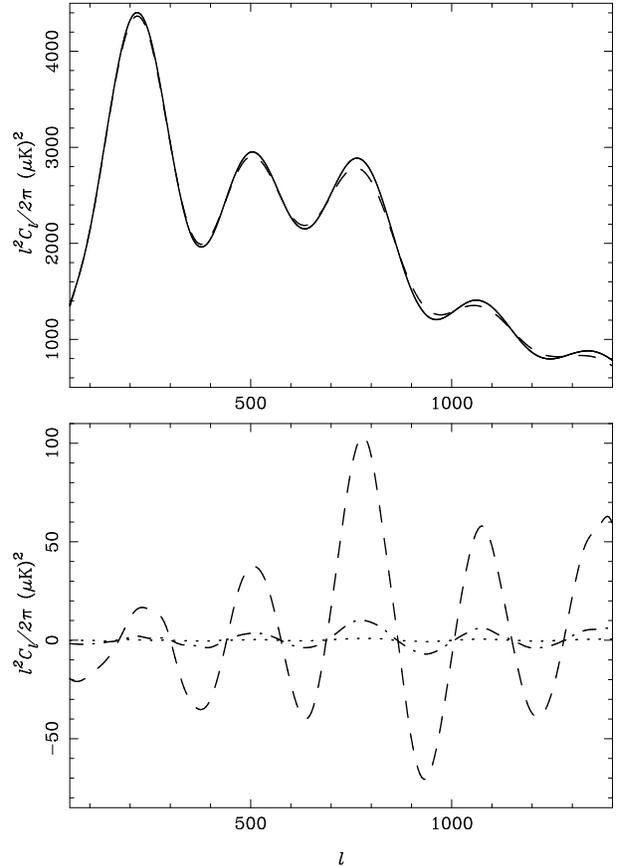,angle=0,
width=8cm}}
\caption{
Effect on the CMB power spectrum of a population of rich cluster of
galaxies.
Upper panel: the solid curve is the unperturbed power spectrum while the
dashed curve represents the `lensed' spectrum if $400$ clusters are
present in the whole sky.
Lower panel: the curves are the difference between the predicted
unperturbed power spectrum and the `lensed' spectrum for three different
cluster abundances corresponding to $400$ (dashed curve), $40$ (dashed-dot
curve) and $4$ (dotted curve) clusters in the whole sky.}
\label{figpower}
\end{figure}

The presence of $4$ clusters in the whole sky corresponds roughly to one
cluster observed in every $100\times 100$-degree field on the sky, while
$40$ clusters correspond to one cluster every $31\times 31$-degree field
and $400$ to one cluster every $10\times 10$-degree field.
We performed, for these 3 cases, Monte-Carlo simulations of the
gravitational effect of the clusters on realisations of primordial
CMB fluctuations, and measured the corresponding power spectra.
Averaging these measured spectra, we obtain an estimate of
the ensemble average power spectrum of CMB fluctuations in the
presence of each proposed cluster population. The results are shown
in Fig.~\ref{figpower}, where they are compared to the ensemble average
power spectrum predicted in the absence of clusters.
We note that our cluster formation model has been developed for
$\Omega_0=1$ and $\Omega_{\Lambda}=0$, while abundances have been
estimated for $\Omega_0+\Omega_{\Lambda}=1$
(see Section~\ref{abundances}).
We will investigate the effect of on our model of $\Omega_0\neq 1$ and
$\Omega_{\Lambda}\neq 0$ in future work.

It is clear that the effect is negligible if only $4$ clusters are present
in the sky.
However, for abundances where $40$ or $400$ clusters are present in the
whole sky, which correspond to $\Omega_0\sim0.2$, the effect is more
pronounced and results in smoothing out of the Doppler peaks in the
inflationary power spectrum.
Therefore, the possibility of such an effect
should be taken into account when determining cosmological parameters
from future CMB observations.
These results are in reasonable agreement with the weak lensing analytical
calculations of Seljak (1996) or Mart\'{\i}nez-Gonz\'alez \& Sanz (1997).

\section{Conclusions}

We apply a new model for the formation of nonlinear cosmic
structures (Paper~I) to the collapse of spherical clusters
of galaxies. The external, expanding Universe and the collapsing
cluster are governed by the same pressureless fluid equations.
These equations are
exact general-relativistic solutions of Einstein equations and no
approximations have been made.
The initial
conditions for the fluid at an early epoch ($z=1000$)
are very simple: we impose a {\it finite} perturbation on the fluid
velocity field that is determined by only three parameters and the
corresponding density perturbation
is inferred assuming that the perturbation arose from primordial
fluctuations.

In Section~\ref{clusterform} we studied the 
formation of a cluster at a redshift
$z=0.08$ and found that density profile and mass distribution
of the resulting cluster are realistic. We also 
computed several characteristic quantities for the cluster, 
such as the total mass contained
within spheres of various radii; the ratio of the central density to that
of the external Universe ($\delta$); and the turnaround radius
($R_t$). Comparing our results with previous authors (e.g. Panek 1992;
Quillis 1995), we find reasonable agreement.

Since photon paths are also easily calculated in our model, 
in Section~\ref{cmbanisot} we studied the gravitational effect of the 
collapsing cluster on CMB photons (i.e. the Rees-Sciama effect).
For a photon traversing the centre of the cluster, we found a
central temperature decrement $\Delta T/T\sim-1\times 10^{-5}$ which is
in reasonable agreement with previous estimates.

Since our model is most relevant to clusters with large infall
velocities, in Section~\ref{highzclusters} we apply
it to clusters with a redshift of $z\ge 1$.
Indeed, such high-redshift clusters are more likely
to be in a state of free-fall collapse than the low-redshift clusters
considered in Sections~\ref{clusterform} and \ref{cmbanisot}.
In particular, we use our model in an attempt to describe the
microwave decrement reported towards the quasar pair
PC1643+4631 A\&B (Jones et al. 1997). Since the quasar pair is possibly
lensed (Saunders et al. 1997), we investigated in Section~\ref{sec_lens}
the lensing properties of a cluster which may explain the observations.
We find that for such a cluster lensing occurs out
to large projected angles $\theta$ from its centre, with an appreciable 
effect still visible at $\theta = 2$ or 3 degrees (see Fig.~\ref{fig8}).

In Section~\ref{sec_cmb} we consider the relative contributions of the
Rees-Sciama and thermal SZ effects to the microwave decrement observed
towards PC1643+4631 A\&B, and show
that the Rees-Sciama effect might contribute significantly
for clusters that can simultaneously produce the required lensing
properties discussed by Saunders et al. (1997).
At $z=1$, such a cluster would have a typical 
central number density of $10^4~{\rm m^{-3}}$
and a core radius of $R_c=0.45~{\rm Mpc}$ (52~arcsec).
The total mass contained within a sphere of radius 2~Mpc is then
$7.8\times 10^{15}~{\rm M_\odot}$.
We also find, however, that in this scenario the time delay between
the light paths for the two quasars images PC1643+4631 A\&B is 
approximately 150 years, as measured in the frame of the (lensed)
quasar. This period might be rather short to explain the slight redshift
difference between the two quasar images. Following very recent
ROSAT observations of PC1643+4631 A\&B by Kneissl (1997), which
suggest that any intervening cluster should be at an even greater
redshift, we also repeat our calculations for a similar cluster at
a redshift $z=2$. 
We find a typical core radius of $R_c=0.74~{\rm Mpc}$ (90~arcsec)
for a central number density of $10^4~{\rm m^{-3}}$.
The total mass contained within a
sphere of radius 2~Mpc is $1.74\times 10^{16}~{\rm M_\odot}$.

Finally, in Section~\ref{cmbpowerspec}, we consider the effect
on primordial microwave background fluctuations of a population of
massive clusters, such as that described in Section~\ref{highzclusters}.
We find that in the case of cluster abundances corresponding to a
$\Omega_0\sim0.2$ and $\Omega_0+\Omega_{\Lambda}=1$ cosmological model,
the Doppler peaks of the CMB power spectrum are slightly smoothed out
by the lensing effects
(see Fig.~\ref{figpower}), confirming weak lensing the results
in Seljak (1996) and suggesting that this effect should be taken
into account when determining cosmological parameters.

\section*{Acknowledgments}

MPH would like to thank Trinity Hall, Cambridge for support in the
form of a Research Fellowship. We also thank the anonymous
referee for numerous and useful comments.

\bsp  
\label{lastpage}
\end{document}